\documentstyle[psfig]{mn}

\newif\ifAMStwofonts

\newcommand{\lapp}{\mbox{\raisebox{-0.3em}{$\stackrel{\textstyle <}{\sim}$}}}
\newcommand{\gapp}{\mbox{\raisebox{-0.3em}{$\stackrel{\textstyle >}{\sim}$}}}

\title[Low-frequency GMRT observations of GRSs]{A multifrequency study of giant radio sources \\
I. Low-frequency Giant Metrewave Radio Telescope observations of selected sources}
\author[C. Konar et al.]
       {C. Konar$^{1,2}$$\thanks{E-mail: chiranjib@iucaa.ernet.in (CK), jamrozy@oa.uj.edu.pl (MJ), 
         djs@ncra.tifr.res.in (DJS), machalsk@oa.uj.edu.pl (JM)}$, M. Jamrozy$^{3}$, 
        D. J. Saikia$^{1}$ and J. Machalski$^{3}$ \\
$^{1}$ National Centre for Radio Astrophysics, TIFR, Pune University Campus, Post Bag 3,
Pune 411 007, India \\
$^{2}$ Inter-University Centre for Astronomy and Astrophysics, Pune University Campus, Post Bag 4,
Pune 411 007, India \\
$^{3}$ Obserwatorium Astronomiczne, Uniwersytet Jagiello\'nski, ul. Orla 171, 30244 Krak\'ow, Poland
}
\date{Accepted.    Received }

\pagerange{\pageref{firstpage}--\pageref{lastpage}}
\pubyear{  }

\begin{document}

\maketitle

\label{firstpage}

\begin{abstract}
We present low-frequency observations with the Giant Metrewave Radio Telescope (GMRT) of 
a sample of giant radio sources (GRSs), and high-frequency observations of three of 
these sources with the Very Large Array (VLA). From multifrequency observations
of the lobes we estimate the magnetic field strengths using three
different approaches, and show that these differ at most by a factor of $\sim$3. 
For these large radio sources
the inverse-Compton losses usually dominate over synchrotron losses when estimates of the
classical minimum energy magnetic field are used, consistent with earlier studies.
However, this is often not true if the magnetic fields are close to the values estimated
using the formalism of Beck \& Krause. We also examine the spectral indices of the cores 
and any evidence of recurrent activity in these sources. We probe the environment using
the symmetry parameters of these sources and suggest that their environments are often
asymmetric on scales of $\sim$1 Mpc, consistent with earlier studies. 
\end{abstract}


\section{Introduction}
Giant radio sources (GRSs) are defined to be those which have a projected linear size
$\gapp$1 Mpc (H$_o$=71 km s$^{-1}$ Mpc$^{-1}$, $\Omega_m$=0.27,
$\Omega_{vac}$=0.73, Spergel et al. 2003). These are useful for studying a number 
of astrophysical problems which include probing the late stages of 
evolution of radio sources, constraining orientation-dependent unified schemes
and probing the intergalactic medium at different redshifts
(e.g. Subrahmanyan \& Saripalli 1993; Subrahmanyan, Saripalli \& Hunstead 1996;
Mack et al. 1998; Ishwara-Chandra \& Saikia 1999;
Kaiser \& Alexander 1999; Blundell, Rawlings \& Willott 1999 and references therein; 
Schoenmakers 1999;
Schoenmakers et al. 2000, 2001). In addition, these sources are useful for studying the 
effects of electron energy loss in the lobe plasma 
due to inverse-Compton scattering with the Cosmic Microwave Background Radiation (CMBR) photons 
at different redshifts (e.g. Konar et al. 2004) and spectral as well as dynamical ageing analyses 
to understand the evolution of the sources (e.g. Konar et al. 2006; Machalski et al. 2007).
Although the number of GRSs published in the literature has increased significantly, there is
still a dearth of GRSs with sizes $\gapp$2 Mpc and at redshifts $\gapp$1. The 
largest known radio source
is 3C236 (Strom \& Willis 1980) with a size of approximately 4 Mpc while the highest redshift 
GRS known so far is 4C 39.24 at a redshift of 1.88 (Law-Green et al. 1995).

We have selected a representative, but somewhat heterogeneous sample of 10  
edge-brightened, large radio sources spread over about two decades in luminosity 
and lying between redshifts of $\sim$0.1 and 0.8 for observations with the 
Giant Metrewave Radio Telescope (GMRT) and the Very Large Array (VLA) over a large frequency range.
Seven of the ten sources are from the sample of GRSs studied by 
Machalski, Jamrozy \& Zola (2001) and Machalski et al. (2006, hereinafter referred to as MJZK),
while the remaining three are from our earlier study (Konar et al. 2004). The Machalski et al.
sample was compiled from the  NRAO VLA Sky Survey (NVSS, Condon et al. 1998) and the 
Faint Images of the Radio Sky at Twenty-centimeters (FIRST) survey (Becker,
White \& Helfand 1995). Due to limitations of
observing time we restricted our study to a sample of ten sources.
The objectives of these observations are to estimate their spectral ages, 
injection spectral indices ($\alpha_{\rm{inj}}$), and also examine the range of magnetic field estimates 
using different
formalisms and their effect on the estimates of the spectral ages. It is worth noting here that
a determination of the spectra of the lobes over a large frequency range from our observations
would help get reliable estimates of the physical parameters of the lobes such as their magnetic fields.
We also explore any evidence of recurrent activity and attempt to probe the environments of these sources 
on Mpc scales using their symmetry parameters. 

With the current cosmological parameters, 
eight of the 10 sources in our sample have projected linear sizes $\gapp$1 Mpc.
In this paper (Paper I) we report GMRT observations of nine of these sources
at a number of frequencies ranging from $\sim$240 to 1300 MHz, and also VLA observations
for three of the sources, namely J1313+6937 at 4873 MHz, J1604+3731 at 4860 and 9040 MHz 
and J1702+4217 at 4860 and 8440 MHz.
GMRT images of J1343+3758 at 316 and 604 MHz are presented here, although the source has been studied
earlier by Jamrozy et al. (2005). The observations presented here for this sample provide information on 
their low-frequency emission and help establish the spectra of the lobes over a large frequency  range. 
In this paper these spectra have been used to estimate the magnetic fields of the lobes using three
different formalisms to understand the possible range of these values, 
make a comparative study of the relative importance of inverse Compton and synchrotron losses,
and explore some of the consequences. We also discuss the core strengths and symmetry parameters of 
these sources to probe their environments and compare with earlier studies.

In a forthcoming paper (Jamrozy et al. 2007; hereinafter referred to as Paper II)
we use the results of our observations to 
estimate the spectral ages and injection spectral indices of these sources and examine 
possible correlations of $\alpha_{\rm {inj}}$ with luminosity, redshift and size. 

\begin{table}
\caption{ Observing log }
\begin{tabular}{l c c c c }

\hline
Teles-    & Array  & Obs.   &  Sources               & Obs.       \\
cope      & Conf.  & Freq.  &                        & Date       \\
          &        & MHz    &                        &            \\
  (1)     &  (2)   & (3)    &   (4)                  & (5)        \\
\hline
GMRT      &        & 241    & J1155+4029             & 2005 Mar 16 \\
GMRT      &        & 239    & J1604+3438             & 2005 Dec 28 \\
GMRT      &        & 334    & J0912+3510             & 2003 Sep 20 \\
GMRT      &        & 316    & J0927+3510             & 2003 Sep 20 \\
GMRT      &        & 334    & J1155+4029             & 2005 Dec 25 \\
GMRT      &        & 316    & J1343+3758             & 2003 Sep 20 \\
GMRT      &        & 334    & J1604+3731             & 2006 Mar 09 \\
GMRT      &        & 334    & J1604+3438             & 2006 Apr 04 \\
GMRT      &        & 334    & J2312+1845             & 2005 Sep 08 \\
GMRT      &        & 604    & J0720+2837             & 2003 Sep 05 \\     
GMRT      &        & 606    & J0912+3510             & 2003 Sep 05 \\
GMRT      &        & 606    & J0927+3510             & 2003 Sep 06 \\
GMRT      &        & 604    & J1343+3758             & 2003 Sep 06 \\
GMRT      &        & 605    & J1155+4029             & 2005 Mar 16 \\
GMRT      &        & 614    & J1604+3438             & 2005 Dec 28 \\
GMRT      &        & 613    & J1604+3731             & 2004 Jan 01 \\
GMRT      &        & 602    & J1702+4217             & 2004 Jan 17 \\
GMRT      &        & 1289   & J1604+3731             & 2006 Jan 01 \\
GMRT      &        & 1288   & J1702+4217             & 2005 Dec 04 \\
GMRT      &        & 1258   & J1155+4029             & 2005 Nov 30 \\
GMRT      &        & 1265   & J1604+3438             & 2006 Jul 12 \\
VLA$^a$   &  D     & 4860   & J1604+3731             & 1993 Dec 23 \\
VLA$^a$   &  D     & 4860   & J1702+4217             & 1993 Dec 23 \\
VLA       &  D     & 4873   & J1313+6937             & 2000 Aug 18 \\
VLA$^a$   &  D     & 8440   & J1702+4217             & 1993 Dec 23 \\
VLA$^a$   &  D     & 9040   & J1604+3731             & 1993 Dec 23 \\
\hline
\end{tabular}

$^a$ unpublished archival data from the VLA \\
\end{table}

\section{Observations and analyses} 
Both the GMRT and the VLA observations were made 
in the standard fashion, with each target source 
observations interspersed with observations of the phase calibrator. The primary 
flux density calibrator was any one of 3C48, 3C147 and 3C286 with all flux 
densities being on the scale of Baars et al. (1977). 
The total observing time on the source is about a few hours for the 
GMRT observations while for the VLA observations the time on source
ranges from 10 to 20 minutes. The low-frequency GMRT data were 
sometimes significantly affected by radio frequency interference,
and these data were flagged. All the data were analysed in the standard 
fashion using the NRAO {\tt AIPS} package. All the data were self calibrated 
to produce the best possible images

\begin{figure}
\vbox{
   \psfig{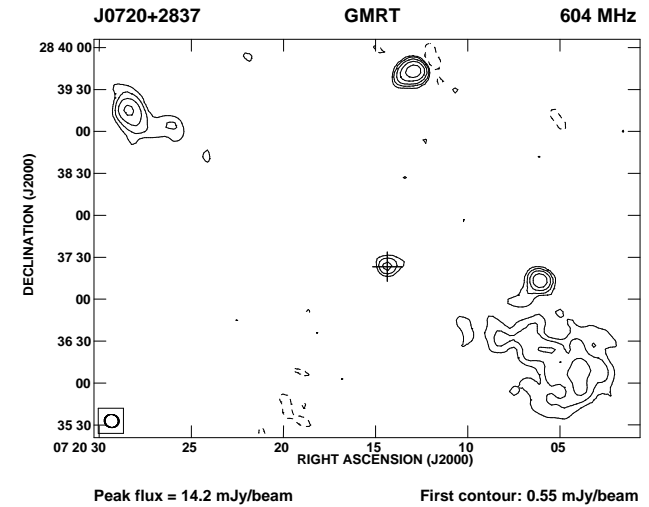}
    }
\caption[]{GMRT image of J0720+2837 at 604 MHz. In this figure as well
as in all the other images of the sources, the peak brightness and
the first contour level are given below each image. The 
contour levels are $-1$, 1, 2, 4, 8 $\ldots$ times the first contour
level. In all the images the restoring beam is indicated by an ellipse 
and the + sign indicates the position of the optical host galaxy. 
           }
\end{figure}

\begin{figure}
\hbox{
     \psfig{file=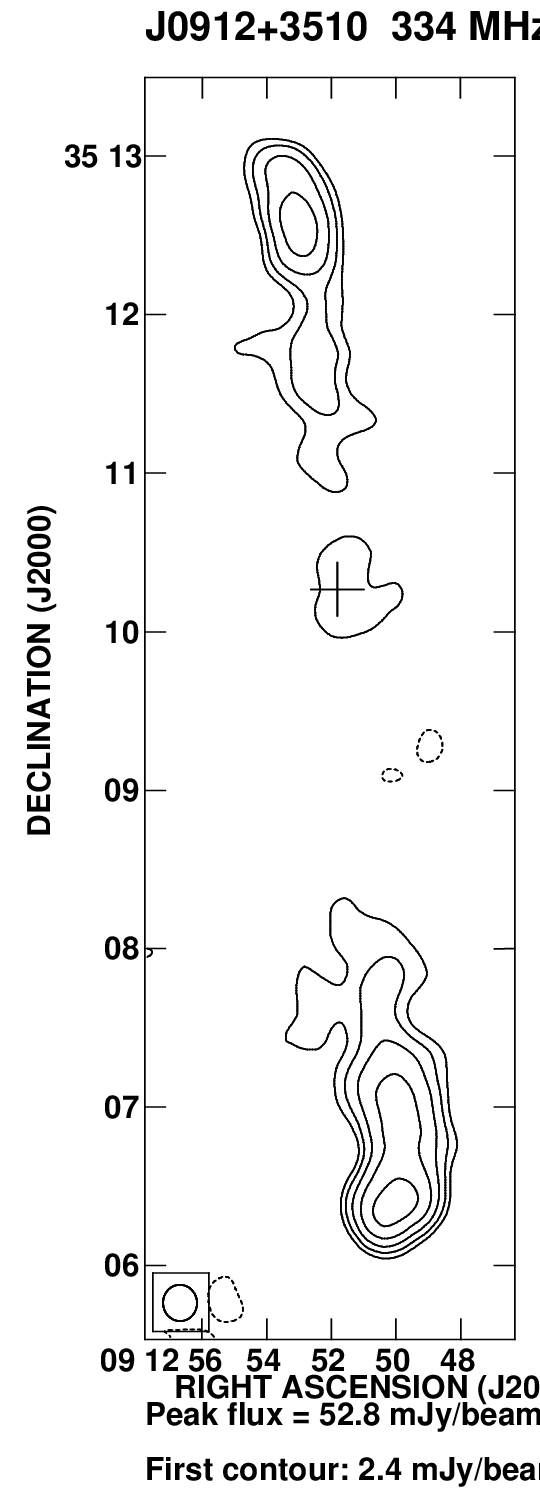,width=1.5in,angle=0}
     \psfig{file=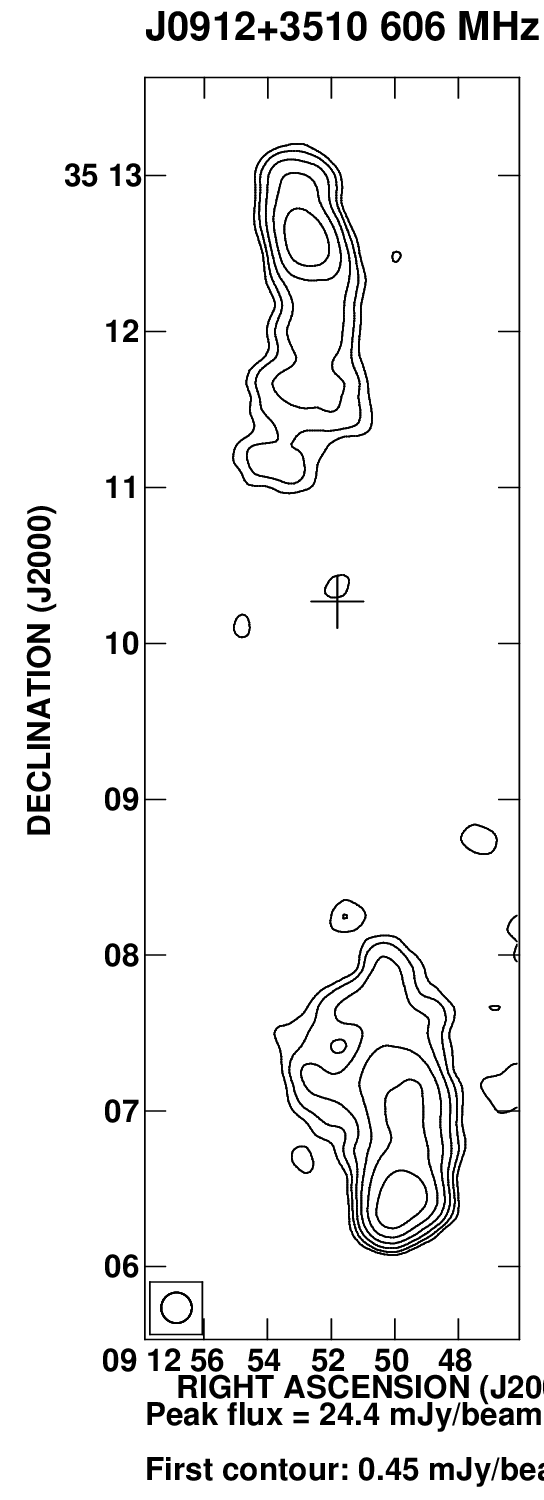,width=1.508in,angle=0}
      }
\contcaption{}
\end{figure}

The observing log for both the GMRT and the VLA observations is given in 
Table 1 which is arranged as follows. Columns 1 and 2 show the name of the 
telescope, and the array configuration for the VLA observations;
column 3 shows the frequency of the observations in MHz, while 
columns 4 and 5 list the sources observed and the dates of the observations 
respectively. 

\begin{figure}
\vbox{
     \psfig{file=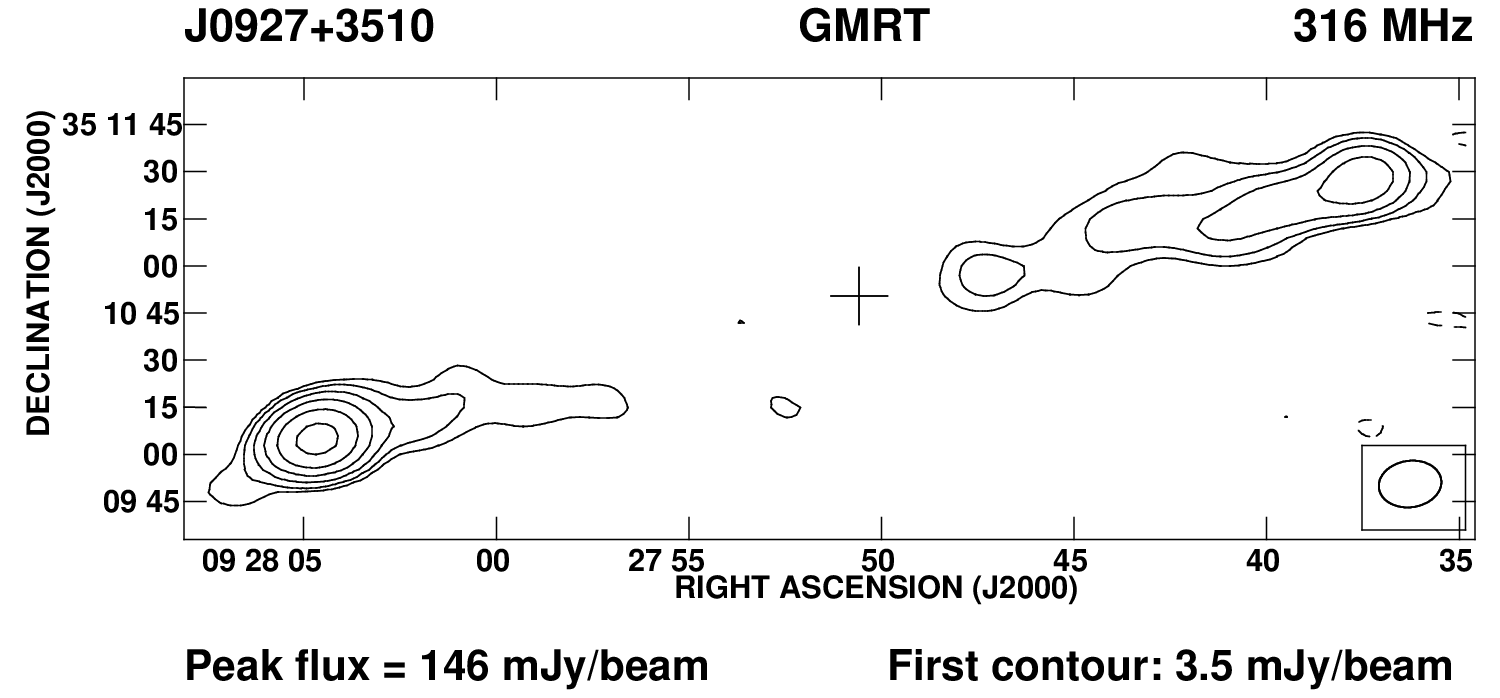,width=3.25in,angle=-90}
     \psfig{file=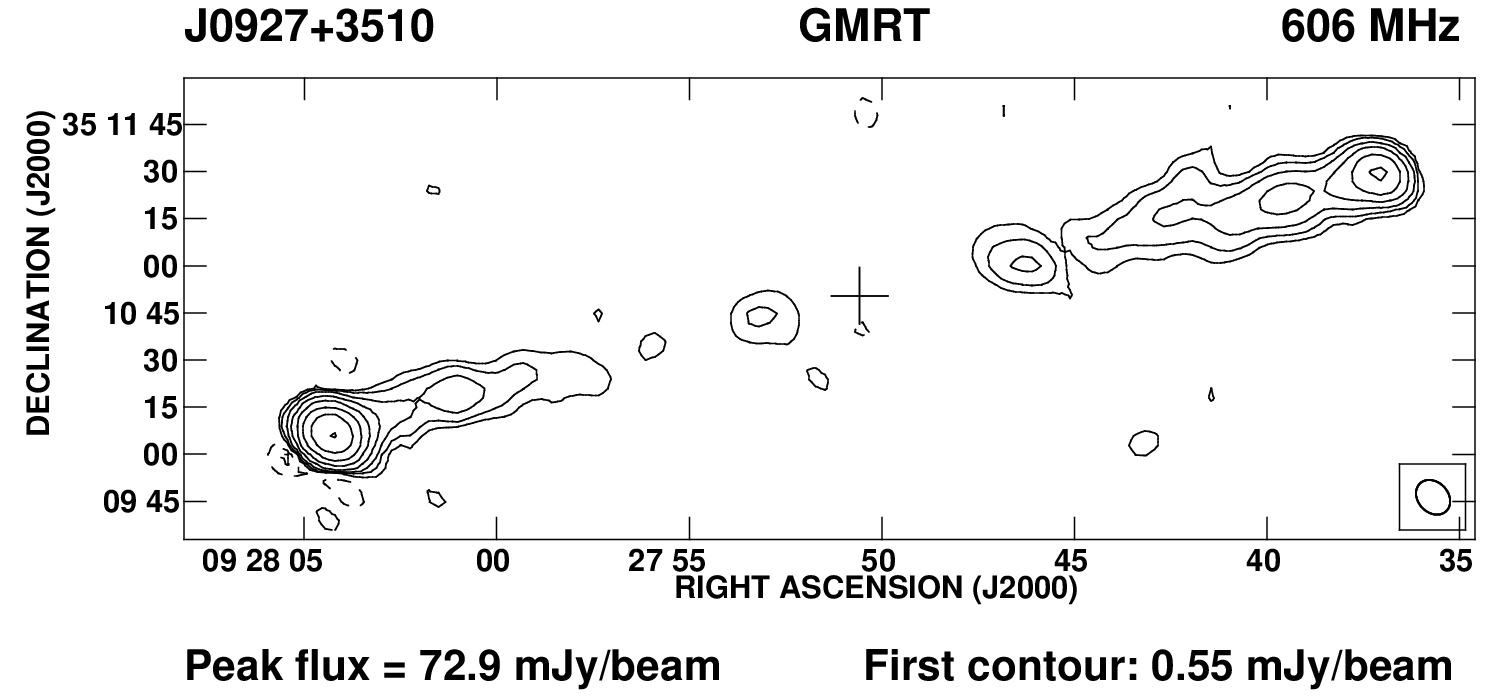,width=3.25in,angle=-90}

    }
\contcaption{}
\end{figure}

\section{Observational results}
The GMRT images of the sources at the different frequencies are presented in
Figure 1, while the observational parameters and some of the observed
properties are presented in Table 2, which is arranged as follows.
Column 1: Name of the source; column 2: frequency  of observations in units of MHz, with
the letter G or V representing either GMRT or VLA observations;
columns 3-5: the major and minor axes of the restoring beam in arcsec and its position angle 
(PA) in degrees;
column 6: the rms noise in units of mJy beam$^{-1}$; column 7: the integrated flux density of the
source in mJy estimated by specifying an area enclosing the entire source. We examined the
change in flux density by specifying different areas and found the difference to be within
a few per cent. The flux densities at different frequencies have been estimated over 
similar areas.  Columns 8, 11 and 14: component designation, where W, E, N, S and C denote 
the western, eastern, northern, southern and core components respectively;
columns 9 and 10, 12 and 13, and 15 and 16: the peak and total flux densities of each of the
components in units of mJy beam$^{-1}$ and mJy respectively. The core flux densities were sometimes 
evaluated by imaging the
source using the longer spacings so that the core appears reasonably isolated. The superscript $g$ 
indicates that the flux densities have been estimated from a two-dimensional Gaussian fit to the 
core component. In this Table we also list the flux densities at 4860 MHz for seven of the sources
whose images have been published by MJZK, and also the flux densities for J1313+6937 at 605 and 
1425 MHz and for J2312+1845 at 1425 and 4866 MHz from the images published by Konar et al. (2004). 

\subsection{Notes on the sources}
Notes on the overall structure and spectra of the cores are
presented in this paper while those based on spectral ages
of the lobes are presented in Paper II.

J0720+2837: The eastern lobe has a more prominent hot-spot compared 
with the western lobe, the ratio of the peak flux density being $\sim$1.7, 
while the western lobe has more extended emission with the ratio of the total 
flux density being $\sim$2.5 in our 604-MHz image. By convolving the 604-MHz image 
(Fig. 1) to a similiar resolution as that of the VLA 4860-MHz image
(MJZK) yields a spectral index, $\alpha$, (S$\propto\nu^{-\alpha}$), 
for the core of 0.63 and 0.59 using the peak and 
total flux densities respectively estimated from two-dimensional Gaussian fits.   
The flux densities of the two unrelated sources in the field, the northern one
at RA: 07$^{\rm h}$ 20$^{\rm m}$ 13.$^{\rm s}$02, Dec: 28$^\circ$ 39$^\prime$ 42.$^{\prime\prime}$58, 
and the western one at RA: 07$^{\rm h}$ 20$^{\rm m}$ 06.$^{\rm s}$09,
Dec 28$^\circ$ 37$^\prime$ 12.$^{\prime\prime}$9 have also been estimated from the 
similar-resolution images.  All positions are in J2000 co-ordinates.
These are 20.1 and 3.3 mJy at 604 and 4860 MHz respectively for the northern
source, the corresponding values for the western source being 11.3 and 0.74 mJy.
The spectral indices for these sources are 0.87 and 1.31 respectively.

\begin{figure*}
\vbox{
  \hbox{
  \psfig{file=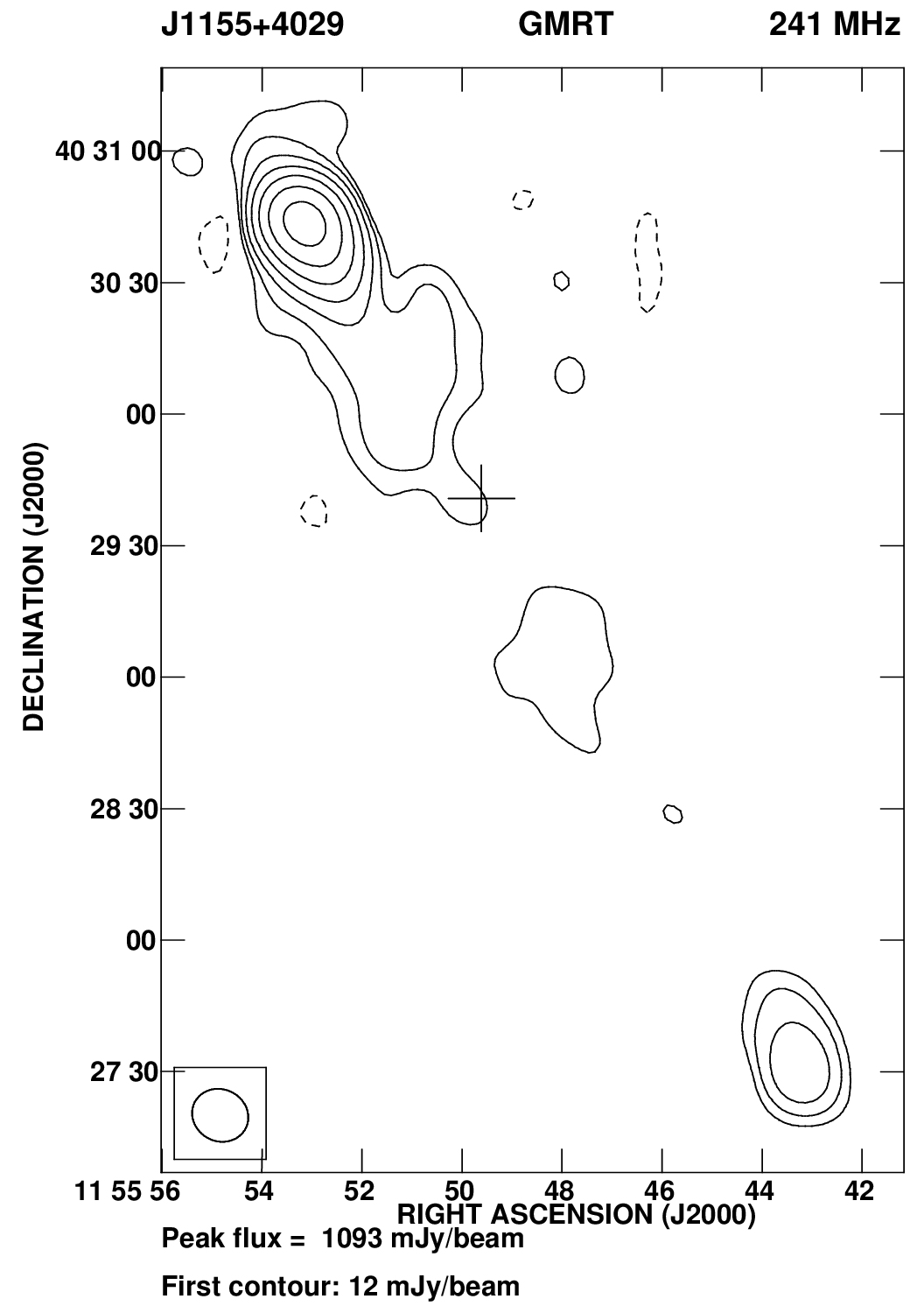,width=3in,angle=0}
  \psfig{file=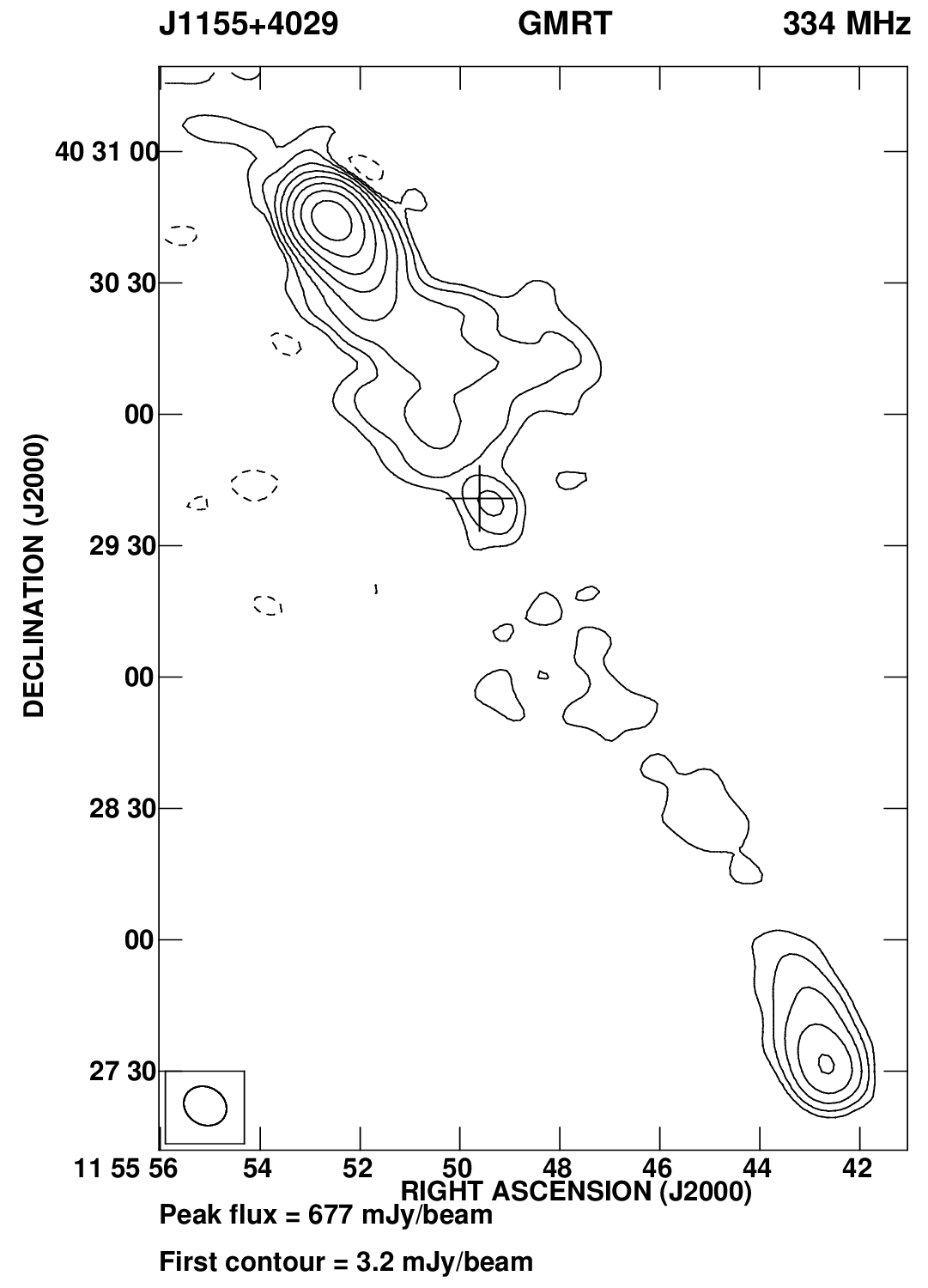,width=3.05in,angle=0}
       }
  \hbox{
  \psfig{file=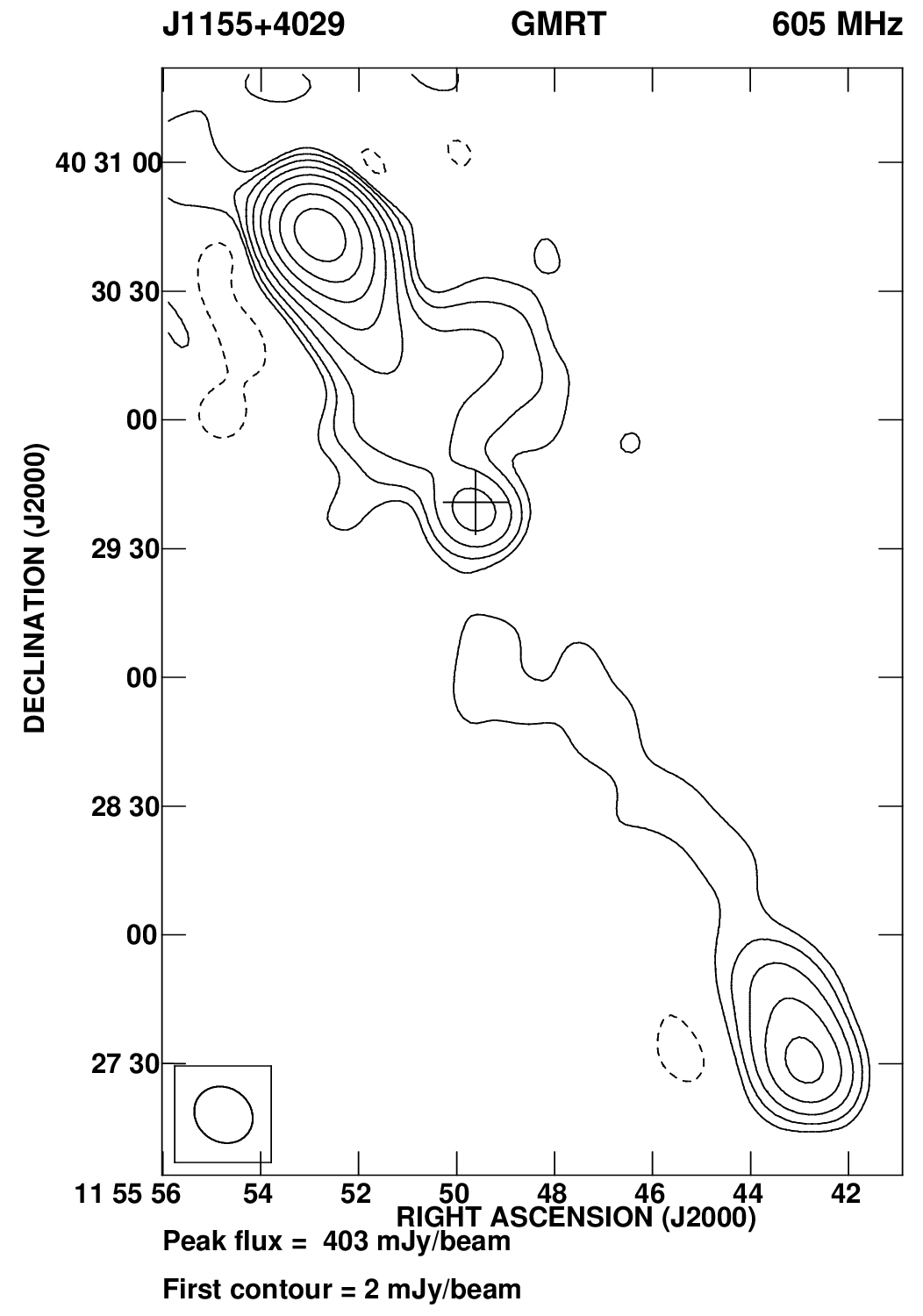,width=3in,angle=0}
  \psfig{file=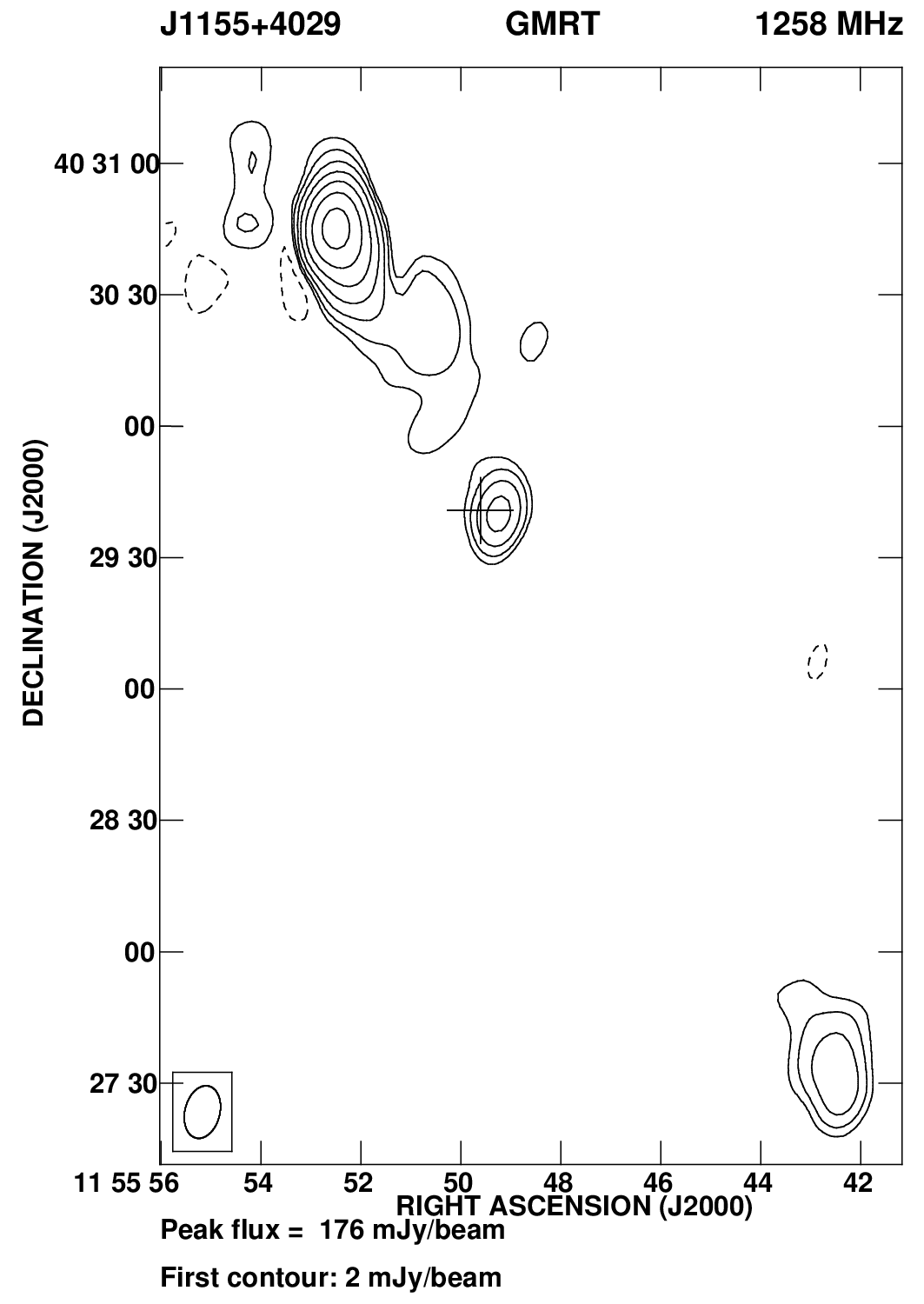,width=3in,angle=0}
       }
}
\contcaption{}
\end{figure*}

J0912+3510: There is a pair of optical galaxies $\sim$2.6 arcsec from the
position of the radio core (MJZK), which has a peak brightness of $\sim$0.4 mJy beam$^{-1}$
in the 4860-MHz image. It is not certain if either of these galaxies might be associated 
with the radio source. The core is not unambigously detected at 606 MHz, the limit being
$\sim$0.6 mJy beam$^{-1}$ from the full-resolution image. Convolving the 606-MHz
image to that of the 4860-MHz one shows that the central component has a peak brightness
of 0.85 mJy beam$^{-1}$, but its position is displaced by $\sim$14 arcsec to the north
of the core seen in the 4860-MHz image.

J0927+3510: This source has been identified as a possible 
DDRG by MJZK based on their 4860-MHz VLA image. Our 606-MHz GMRT
image also shows clearly the inner components with the optical host galaxy in 
between. Using the integrated flux densities at 316, 606 and 4860 MHz from 
images of the same resolution yields spectral indices of $\sim$1.35 and 1.17
for the inner western and eastern components respectively (see also Paper II).
The steep spectral indices suggest that these features are likely to be 
enhancements in the tail emission rather than a new pair of lobes.  

J1155+4029: It is very asymmetric in its arm-length as well as flux density ratios.
The arm-length ratio is $\sim$2.1, while the ratios of the peak and total flux densities
estimated from the 605-MHz image are $\sim$14 and 7 respectively with the nearer (eastern)
lobe being brighter. The ratios are similar at the other frequencies.  In our sample of 10 GRSs,
it has the highest value of core fraction of 0.14 at a frequency of 8 GHz in its rest frame.
However, the core has a spectrum which is peaked at $\sim$1 GHz (see Section 4.2), and it would
be interesting to determine from high-resolution observations whether it might be a  
mas-scale double.  

J1343+3758: This source, which has been studied in some detail by Jamrozy et al. (2005),
has an arm-length ratio of $\sim$1.3 with the more relaxed (western) lobe being closer to the nucleus.
This source has a core with a flat spectrum with $\alpha\sim$0.1  between 604 and 4860 MHz.

\begin{figure}
    \psfig{file=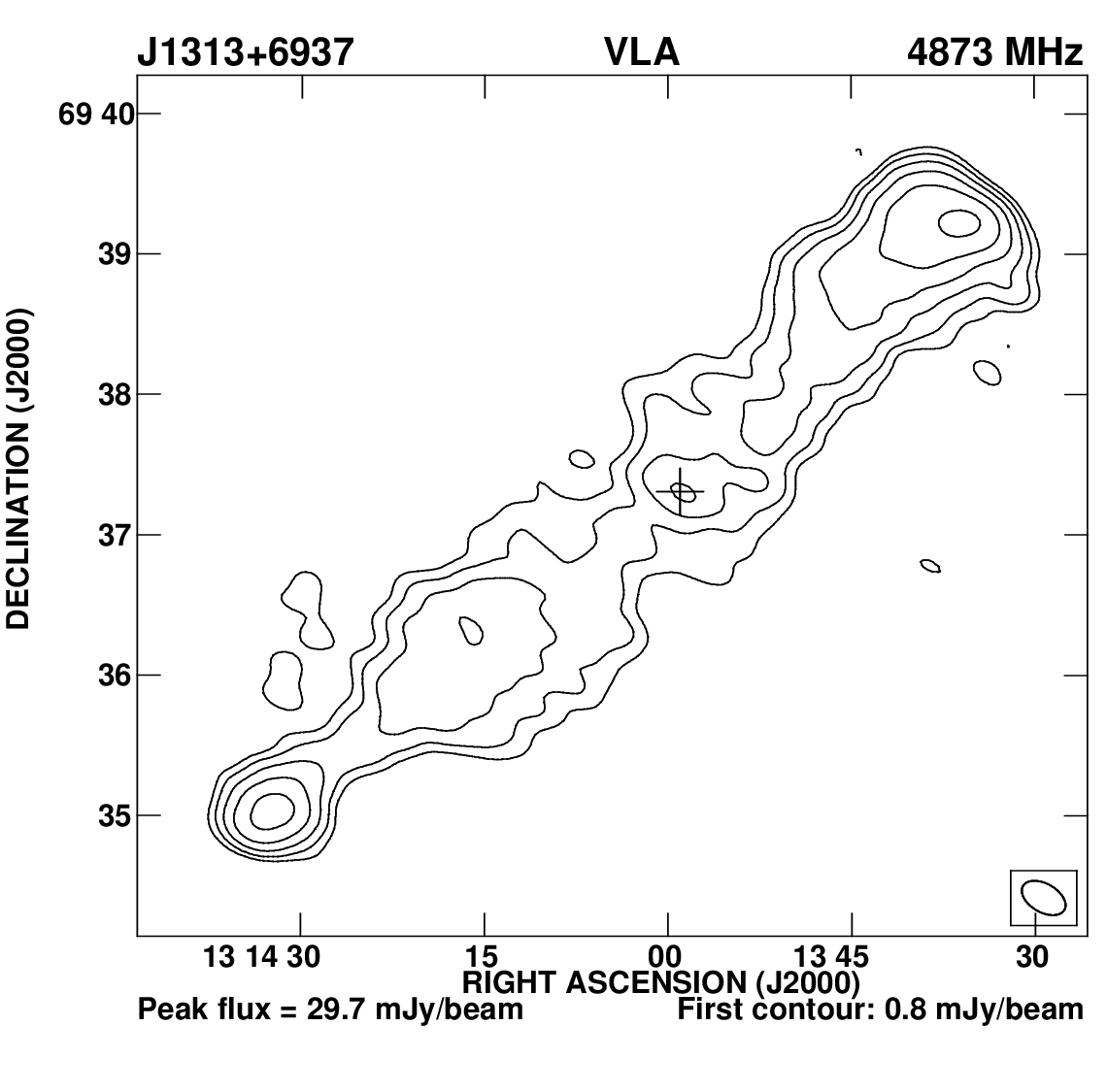,width=3.2in,angle=-90}
\contcaption{}
\end{figure}

\begin{figure}
\vbox{
  \psfig{file=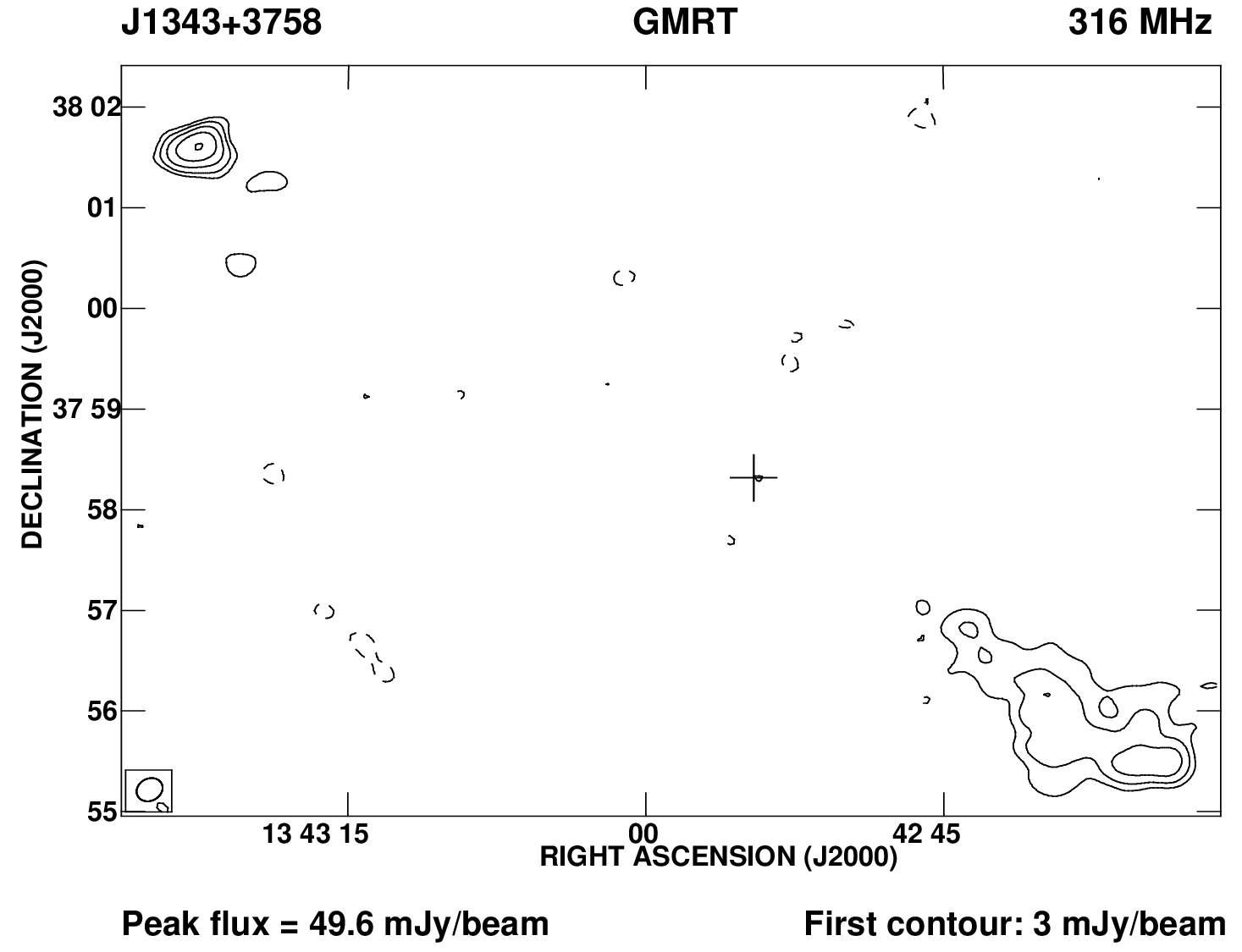,width=3.25in,angle=-90}
  \psfig{file=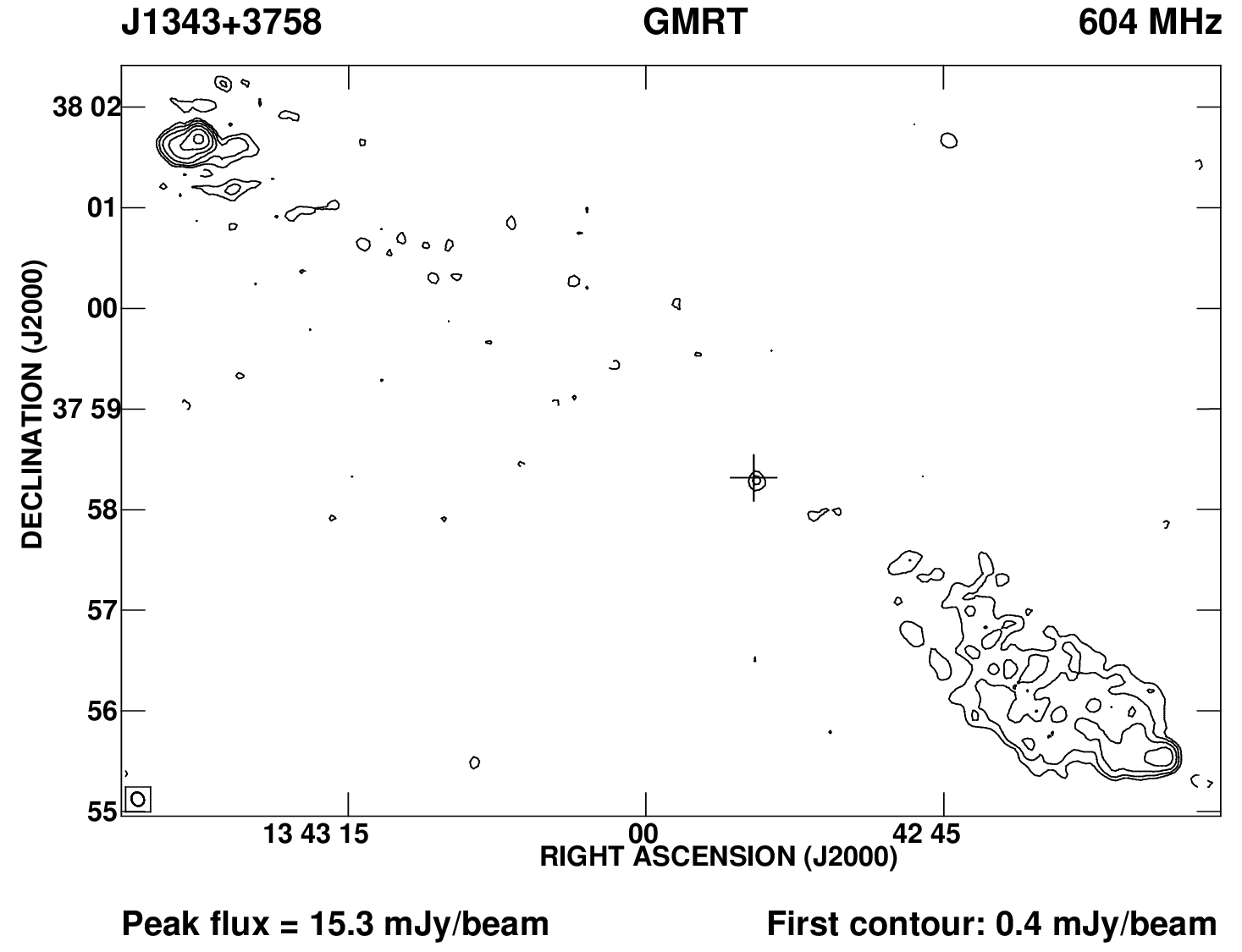,width=3.25in,angle=-90}
     }
\contcaption{}
\end{figure}

\begin{figure*}
\vbox{

  \hbox{
  \psfig{file=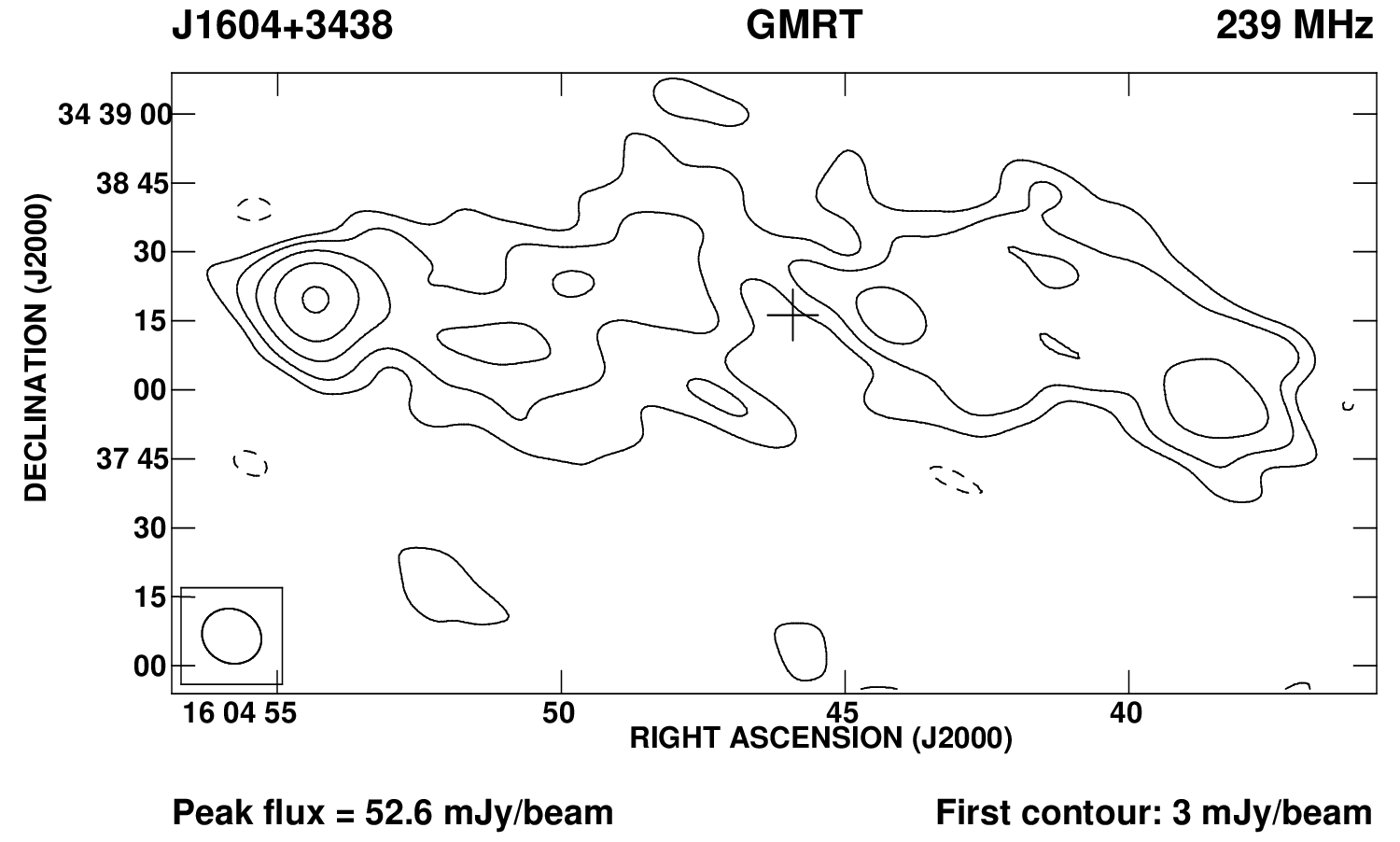,width=3.5in,angle=-90}
  \psfig{file=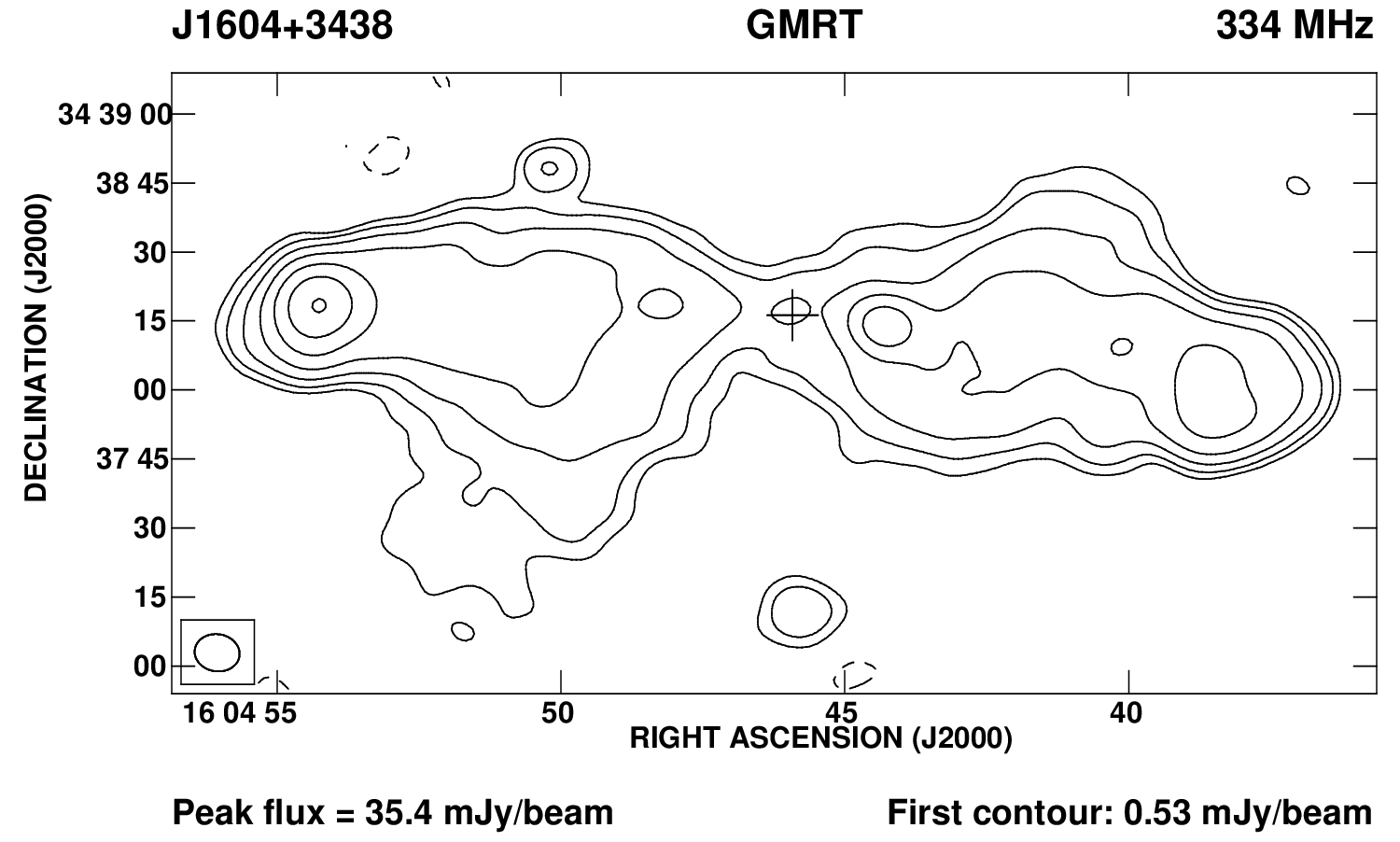,width=3.5in,angle=-90}
  }

 \hbox{
 \psfig{file=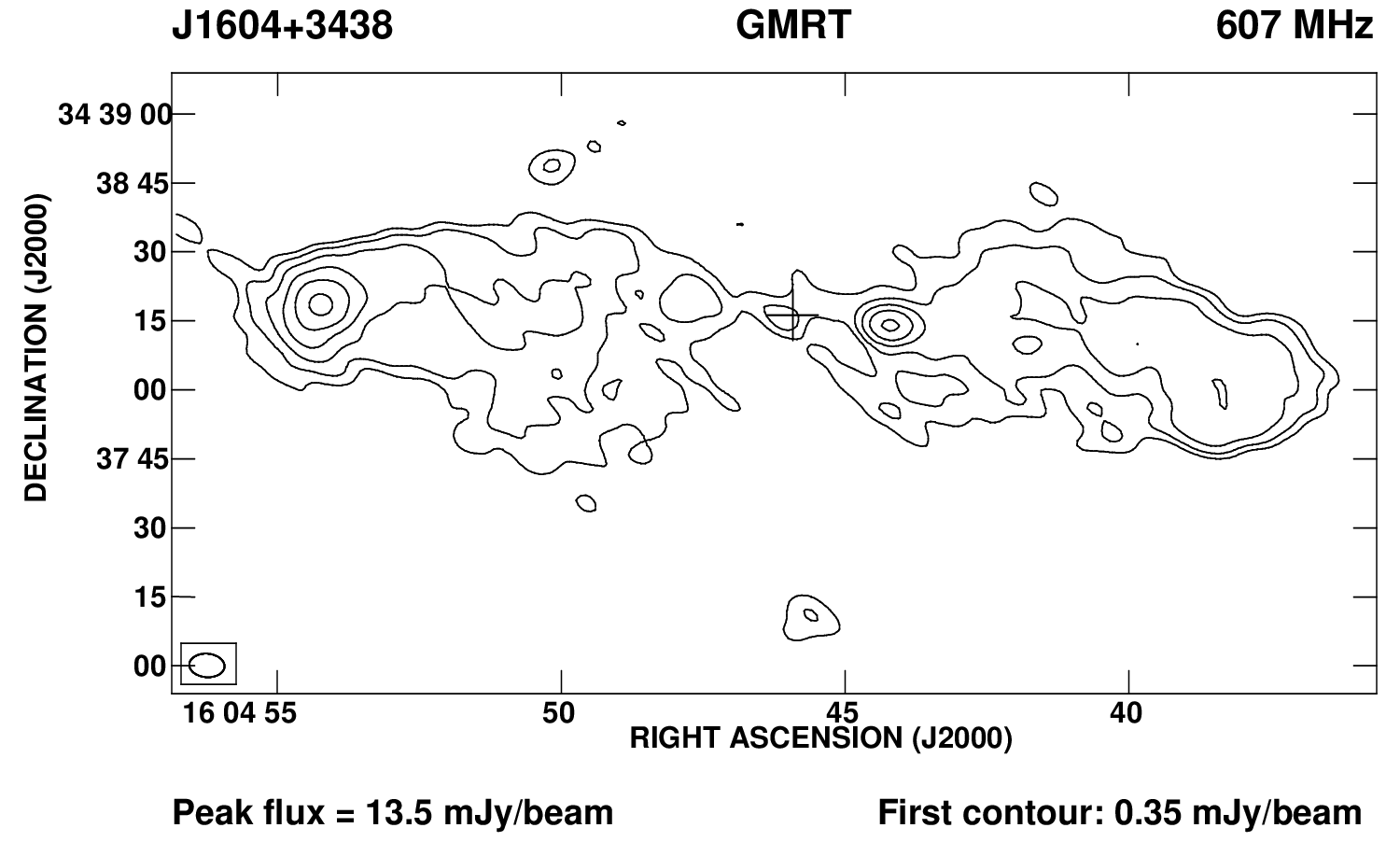,width=3.5in,angle=-90}
 \psfig{file=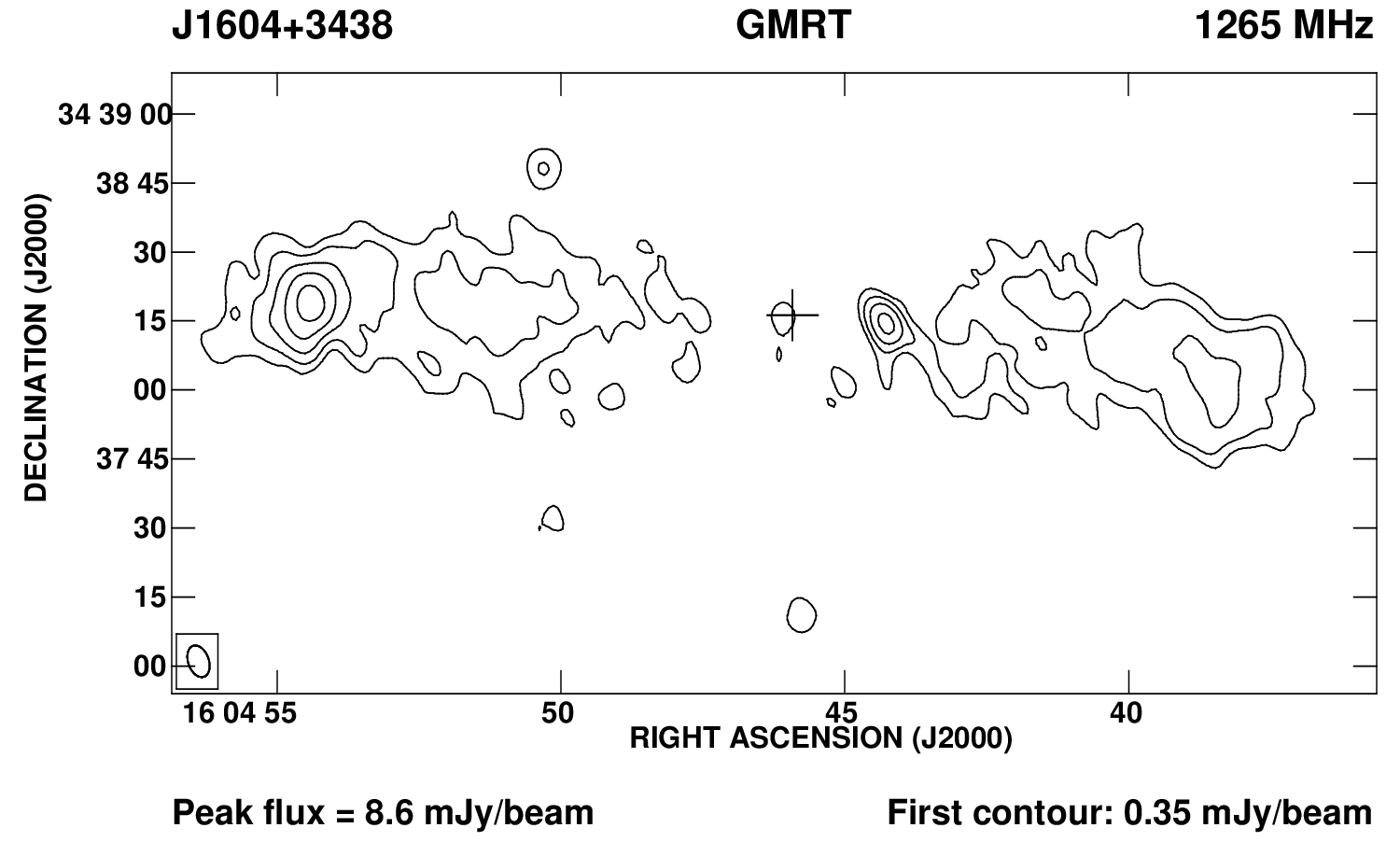,width=3.5in,angle=-90}
}

}
\contcaption{}
\end{figure*}

\begin{table*}
\caption{ The observational parameters and observed properties of the sources}

\begin{tabular}{l l rrr r r r rr l rr r rr}
\hline
Source   & Freq.       & \multicolumn{3}{c}{Beam size}                    & rms      & S$_I$   & Cp  & S$_p$  & S$_t$  & Cp   & S$_p$ & S$_t$ & Cp  & S$_p$   & S$_t$     \\

           & MHz         & $^{\prime\prime}$ & $^{\prime\prime}$ & $^\circ$ &    mJy   & mJy     &     & mJy    & mJy    &      & mJy   & mJy   &     & mJy     & mJy       \\
           &             &                   &                   &          & beam$^{-1}$&       &     &beam$^{-1}$   &        &      & beam$^{-1}$    &       &     & beam$^{-1}$      &           \\ 
   (1)     & (2)   & (3)  & (4)  & (5)  & (6)  & (7)  &(8)& (9)  & (10) & (11)  &   (12)  &(13)  &(14)& (15) & (16)  \\
\hline
J0720+2837 & G604  & 10.1 &  9.3 & 73.3 & 0.18 & 96   & W & 3.2  & 65  & C$^g$&       2.8 &3.3  & E & 5.3 & 26   \\
           & V4860 & 19.1 & 11.9 & 70   & 0.03 & 16   & W & 1.1  & 6.7 & C$^g$&      0.85 &1.0  & E & 2.0 & 7.9  \\

J0912+3510 & G334  & 13.7 & 12.9 & 4    & 0.78 & 488  & N & 28   & 175 &      &           &     & S & 52  & 309  \\
           & G606  & 11.8 & 11.8 & 0    & 0.11 & 257  & N & 13   &  82 &      &           &     & S & 24  & 171  \\
           & V4860 & 20   & 20   & 0    & 0.04 &  56  & N & 5.6  &  20 & C    &$\lapp$0.4 &     & S & 11  &  36 \\

J0927+3510 & G316  & 20   & 14.8 & 98   & 0.92 & 376  & W & 48   & 160 &      &           &     & E & 149 & 214  \\
           & G606  & 12.2 &  9.6 & 43   & 0.11 & 192  & W & 20   &  74 &      &           &     & E &  74 & 118  \\
           & V4860 & 22   & 11.7 & 73   & 0.03 &  27  & W &  4.5 &  11 & C$^g$&      0.14 &0.14 & E &  13 &  16  \\ 
 
J1155+4029 & G241  & 13.1 & 11.8 &  63  & 3.24 &  2120& NE& 1087 & 1774& C    & 14        &     & SW&  96 &  307  \\
           & G334  & 10.1 &  8.6 &  61  & 1.18 &  1502& NE&  684 & 1241& C$^g$& 15        & 24  & SW&  54 &  219  \\
           & G605  & 14.2 & 12.5 &  53  & 0.73 &   769& NE&  403 &  630& C$^g$& 22        & 28  & SW&  41 &  105  \\
           & G1258 & 12.2 &  8.0 & 166  & 0.46 &   367& NE&  175 &  294& C$^g$& 22        & 23  & SW&  15 &   48  \\
           & V4860 & 17.8 & 12.2 &  92  & 0.04 &    99& NE&   55 &   74& C$^g$& 13        & 14  & SW&  6.6&   11  \\

J1313+6937 & G605  &  8.1 &  4.6 & 151  & 0.20 & 2499 & NW&   39 & 1441& C    &$\lapp$2.0 &     & SE&  27 & 1046  \\
           & V1425 & 14.9 &  6.5 & 121  &  0.10& 1383 & NW&  54  & 797 & C$^g$& 4.6       & 7.0 & SE&  40 &  585  \\
           & V4873 & 20.4 & 12.2 &  60  &  0.13&  431 & NW&  30  & 255 & C$^g$& 4.1       & 4.3 & SE&  18 &  172  \\
          
J1343+3758 & G316  & 16   & 13.2 & 117  & 0.83 & 455  &SW & 22   & 318 & C    &$\lapp$3.0 &     &NE &  48 & 144   \\
           & G604  & 8.4  & 7.3  &  29  & 0.11 & 253  &SW & 5.3  & 165 & C$^g$&       1.0 &1.4  &NE &  15 &  87   \\
           & V4860 & 15   & 15   &   0  & 0.03 &  42  &SW & 3.4  &  25 & C$^g$&       1.1 &1.1  &NE & 6.7 &  16   \\

J1604+3438 & G239  & 13.2 & 11.7 &  63  & 1.51 & 518  & W & 21   & 242 &      &           &     & E &  52 & 277   \\
           & G334  &  9.8 &  8.1 &  81  & 0.15 & 510  & W & 14   & 241 & C    &$\lapp$2.4 &     & E &  35 & 270   \\
           & G614  &  7.7 &  5.1 &  86  & 0.07 & 232  & W & 6.6  & 108 & C    &$\lapp$1.0 &     & E &  14 & 125   \\
           & G1265 &  7.1 &  4.5 &  20  & 0.07 & 152  & W & 3.6  &  67 & C    &$\lapp$0.5 &     & E & 8.6 &  83   \\
           & V4860 & 15.0 & 10.0 &  72  & 0.06 &  36  & W & 2.1  &  16 & C$^g$&       1.0 & 1.1 & E & 6.5 &  19  \\ 
       
J1604+3731 & G334  &10.4  & 7.8  &  68  & 0.35 & 522  & N &  41  & 222 & C$^g$&       10  &12   & S & 106 & 284   \\
           & G613  & 7.8  & 4.8  & 171  & 0.10 &  264 & N &  14  & 111 & C$^g$&       5.4 & 6.4 & S &  53 & 142   \\
           & G1289 & 8.3  & 7.7  &  26  & 0.25 &  144 & N &  12  &  61 & C$^g$&       3.8 & 4   & S &  32 &  81   \\
           & V4860 &13.4  &13.1  &  93  & 0.01 &   29 & N & 4.6  &  11 & C$^g$&       1.3 & 1.6 & S &  10 &  17   \\
           & V9040 & 7.4  & 6.7  &  49  & 0.01 &   13 & N & 1.1  & 4.2 & C$^g$&       0.9 & 0.9 & S & 4.0 & 7.7   \\

J1702+4217 & G602  & 12.3 & 4.4  & 124  & 0.10 &  381 & NE&  30  & 220 & C$^g$&       1.5 & 1.9 & SW&  14 & 158   \\
           & G1288 &  2.7 &  2.3 &  170 & 0.06 &  242 & NE&  7.6 & 141 & C$^g$&       1.3 &  1.2& SW& 3.8 & 102   \\
           & V4860 & 13.7 &  13  &  65  & 0.02 &  48.3& NE&  9.7 &  28 & C$^g$&       1.0 &  1.3& SW& 5.9 &  19   \\
           & V8440 &  8.8 &  7.2 &  63  & 0.01 &  23  & NE&  4.1 &  14 & C$^g$&       1.4 &  1.5& SW& 1.8 &   8   \\
 
J2312+1845 & G334  & 14.1 &  8.7 &  85  & 1.90 & 8131 & NE&  954 & 3510& C    &$\lapp$18  &     & SW& 1089& 4537  \\  
           & V1425 & 14.2 & 12.7 & 179  & 0.26 & 1885 & NE& 311  & 802 & C$^g$& 3.2       & 4.1 & SW& 356 & 1069  \\
           & V4866 & 14.2 & 13.7 & 120  & 0.12 &  538 & NE& 102  & 230 & C$^g$& 2.9       & 3.6 & SW& 123 &  307  \\

\hline
\end{tabular}
\end{table*}

J1604+3438: 
This source was listed as a possible DDRG by MJZK on the basis of their VLA image at 
4860 MHz. Although higher-resolution observations are requried to determine the existence
of any hot-spot in the inner structure, the spectral indices of the features which have
been suggested to be the inner double are steep. Convolving the GMRT 1265-MHz image to the
same resolution as that of the VLA 4860-MHz one, the spectral indices of the western 
component of the inner structure is $\sim$1.0 while for the eastern one $\alpha$$\sim$1.8.
The core spectral index from these images is $\sim$0.3. The steep spectral indices could
be affected by contamination with emission from the backflow. Higher resolution images to
determine the structure and spectral indices are required to confirm the DDRG nature of
the source. 

\begin{figure*}
\vbox{
\hbox{
    \psfig{file=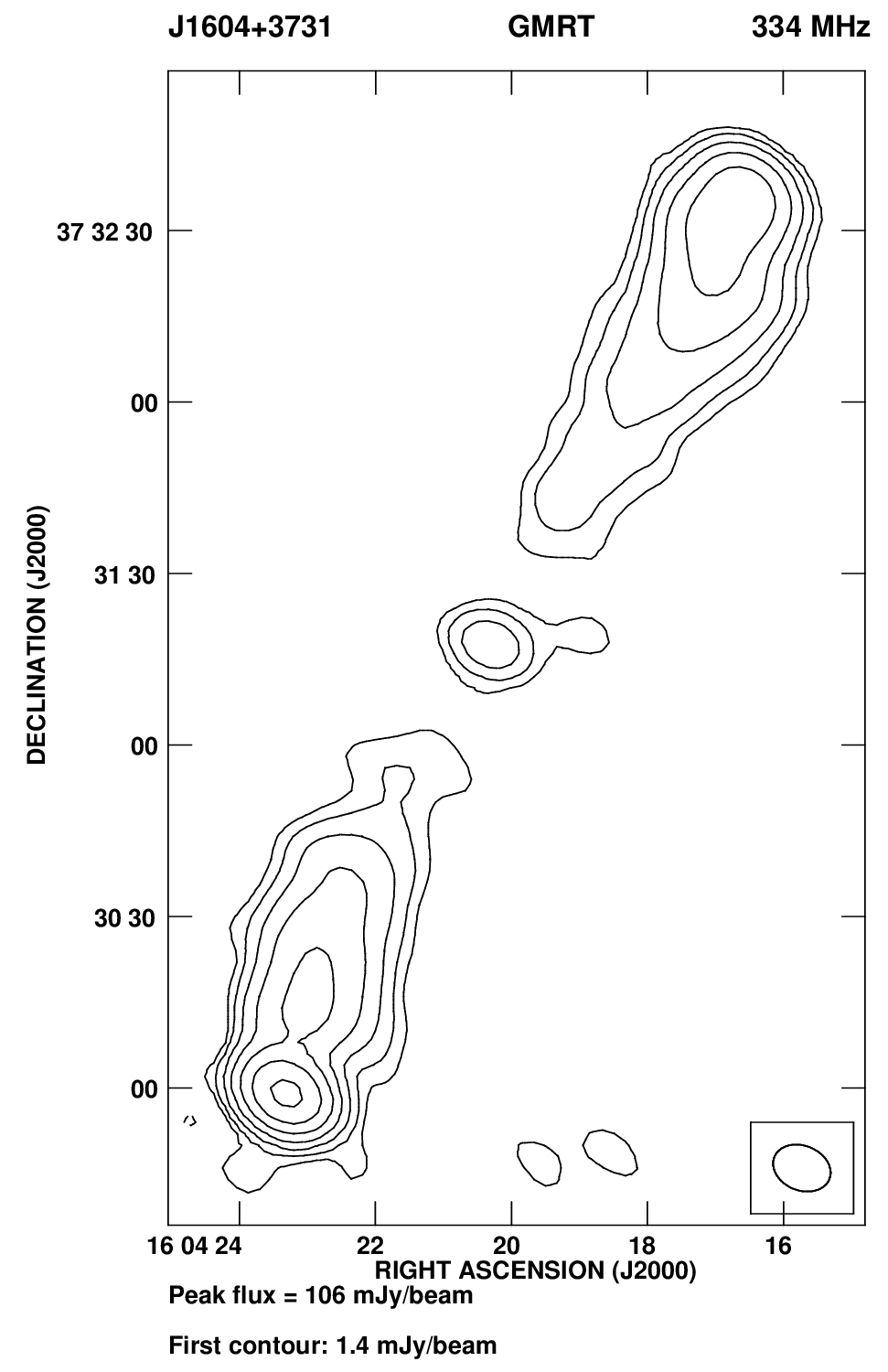,width=3in,angle=0}
    \psfig{file=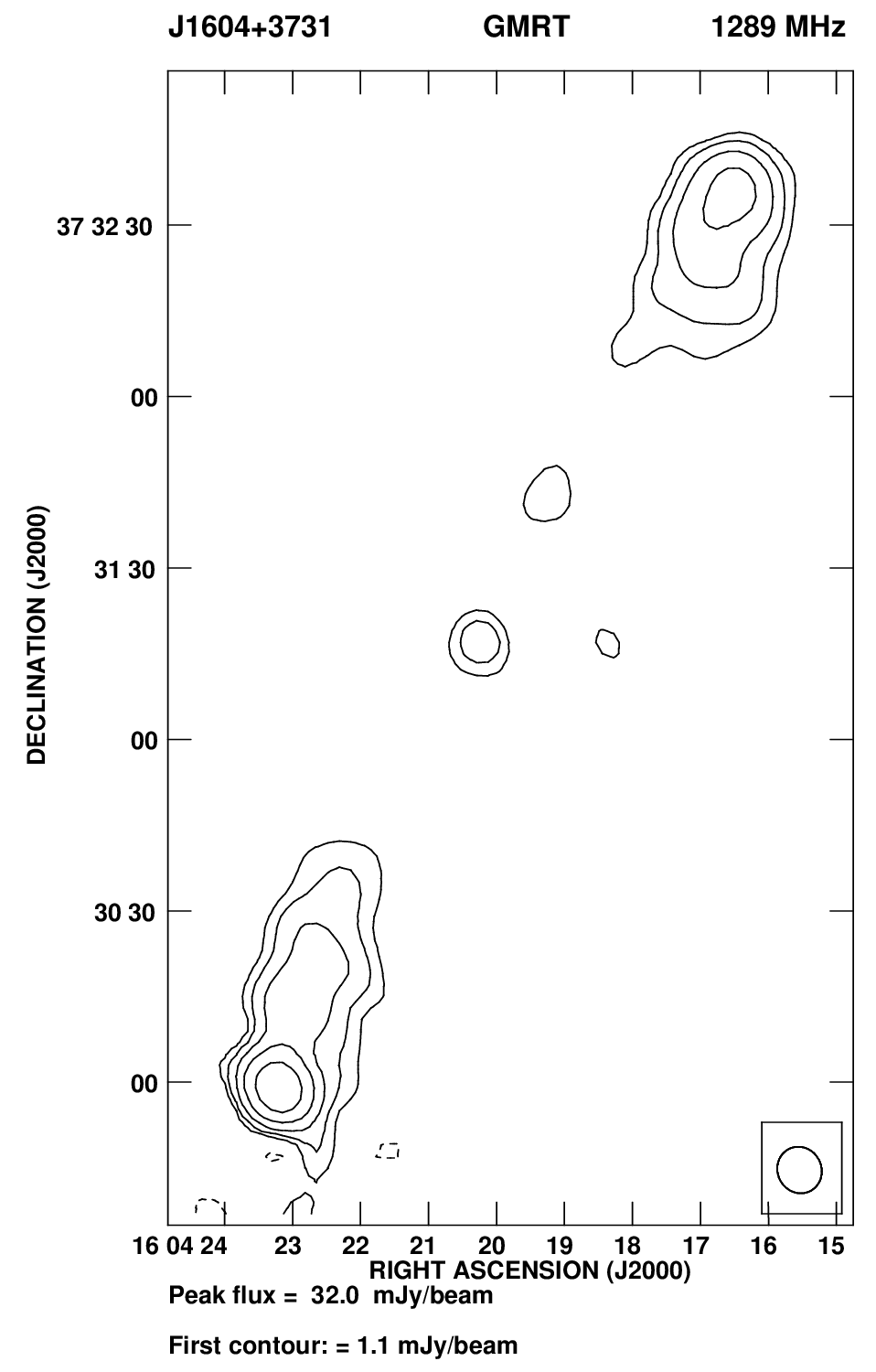,width=3in,angle=0}
     }
\hbox{
   \psfig{file=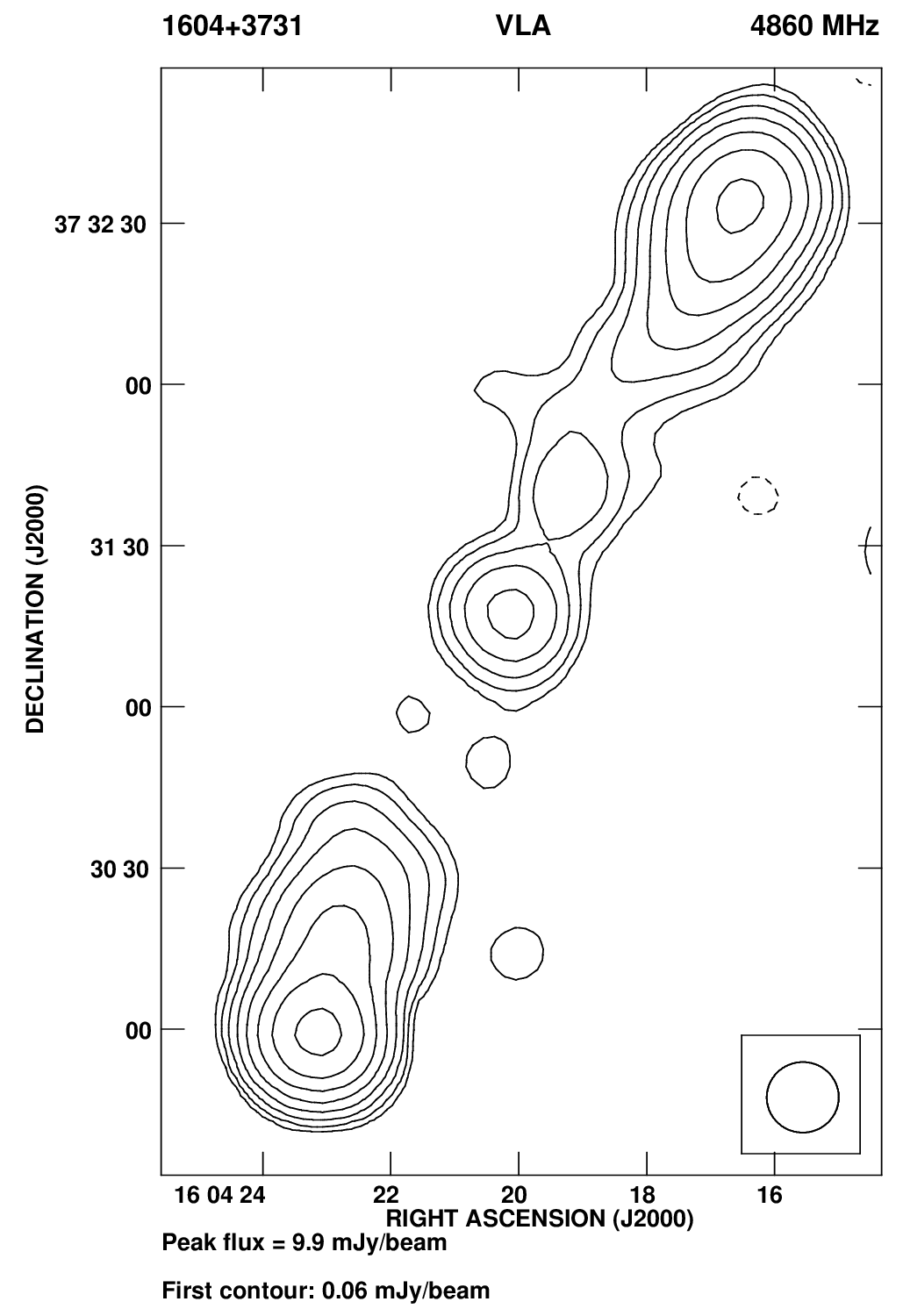,width=3in,angle=0}
   \psfig{file=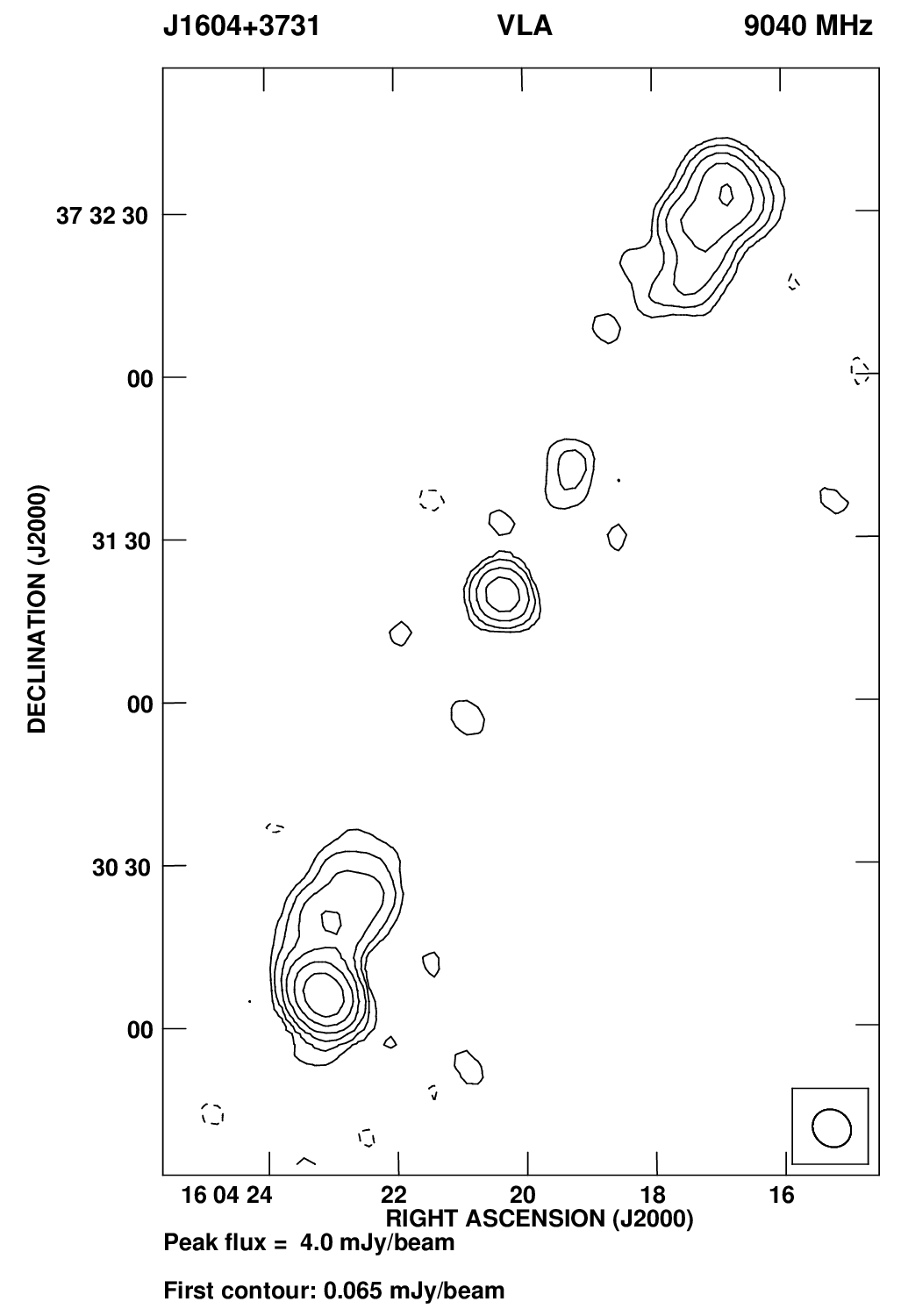,width=3in,angle=0}
    }
}
\contcaption{}
\end{figure*}


\begin{figure*}
\vbox{
    \hbox{
    \psfig{file=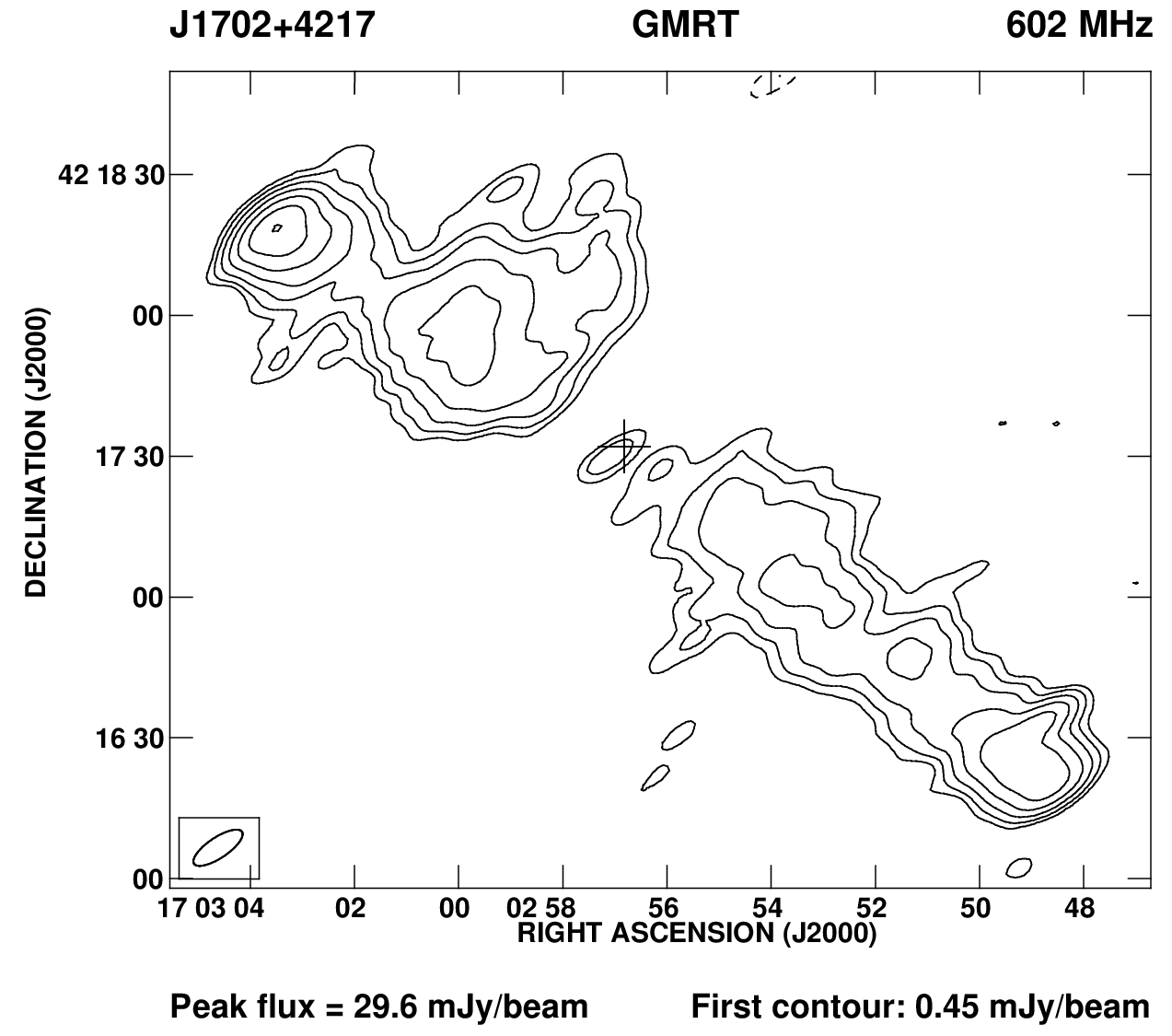,width=3.2in,angle=-90}
    \psfig{file=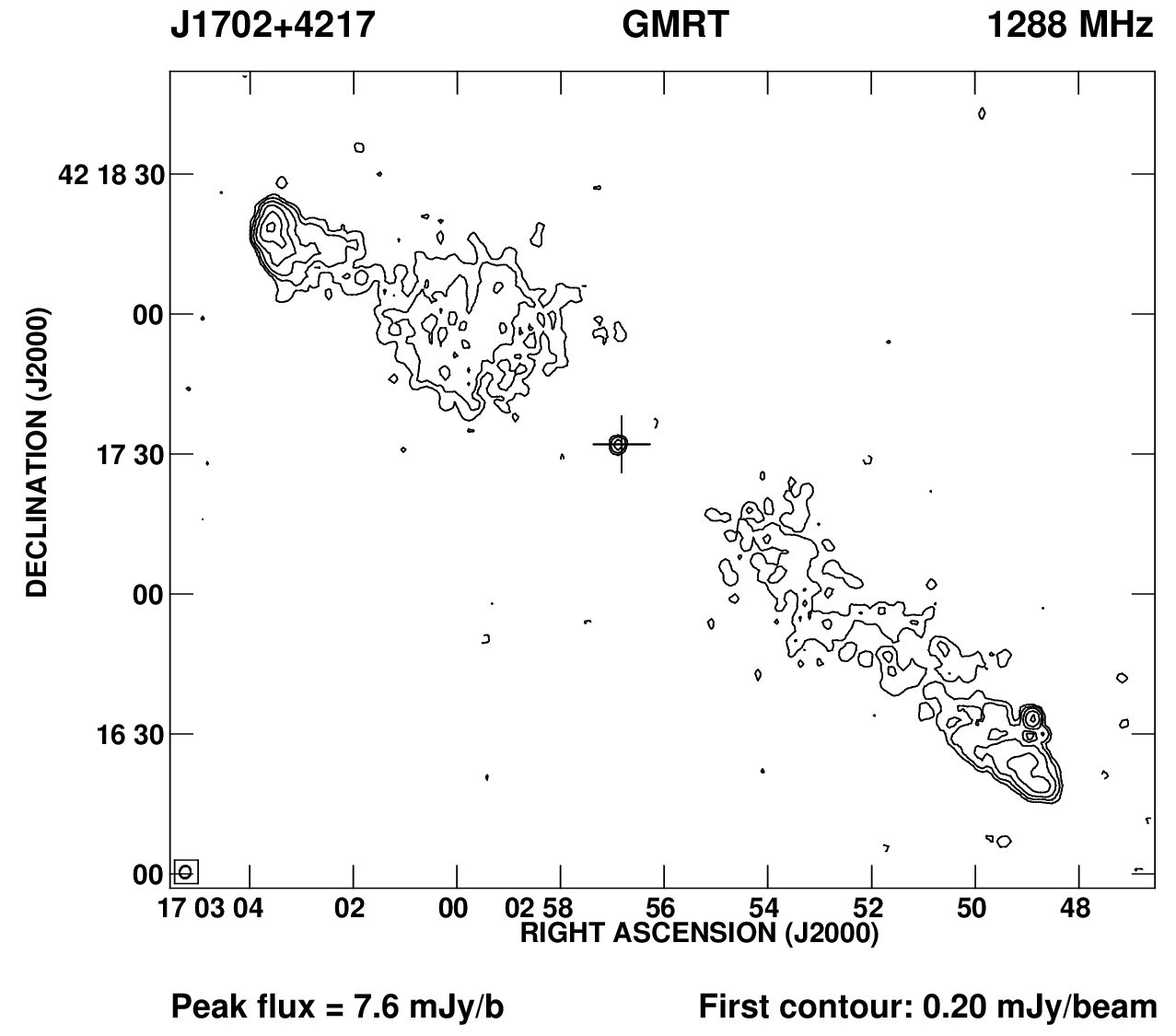,width=3.2in,angle=-90}
         }
    \hbox{
   \psfig{file=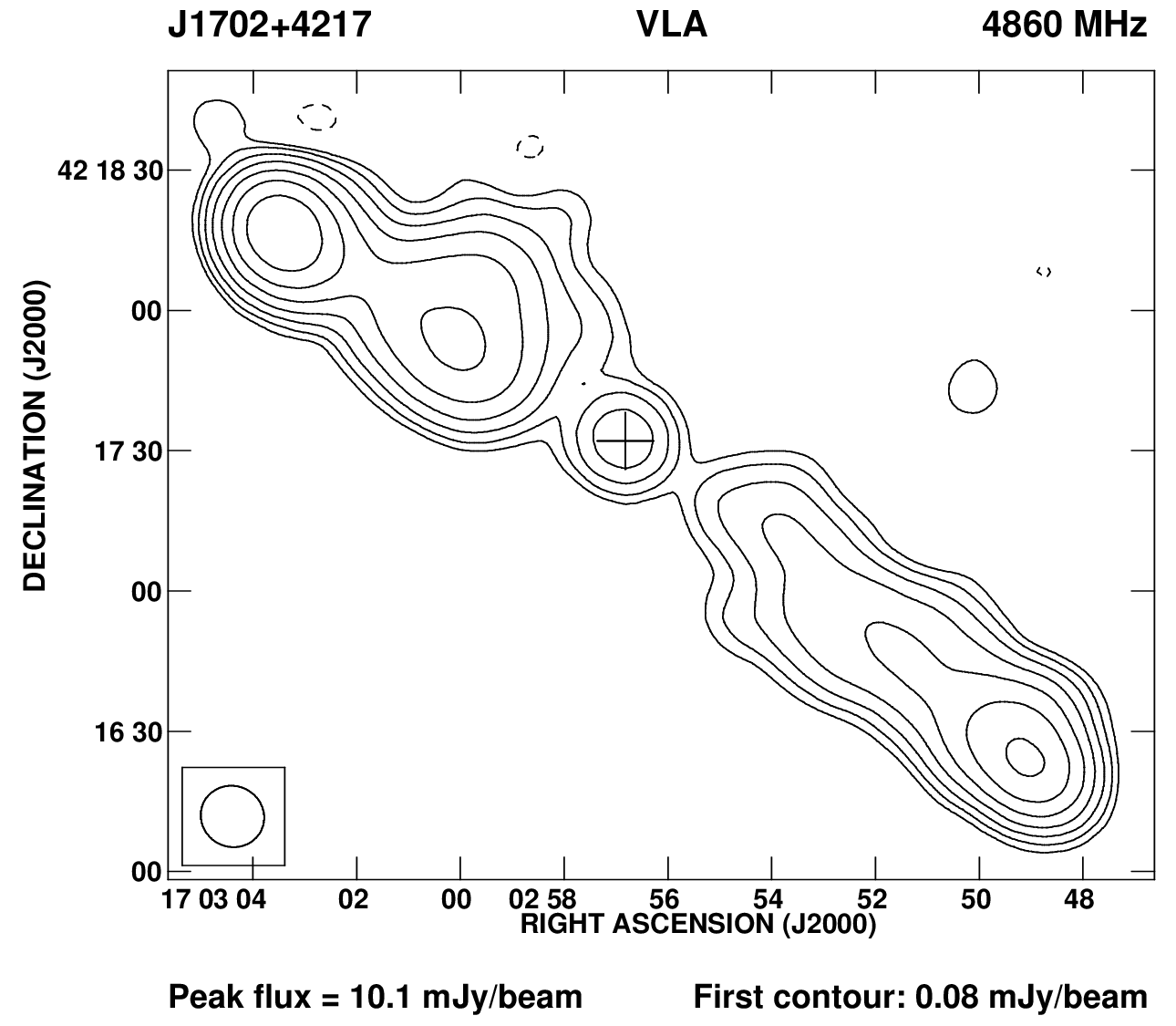,width=3.2in,angle=-90}
   \psfig{file=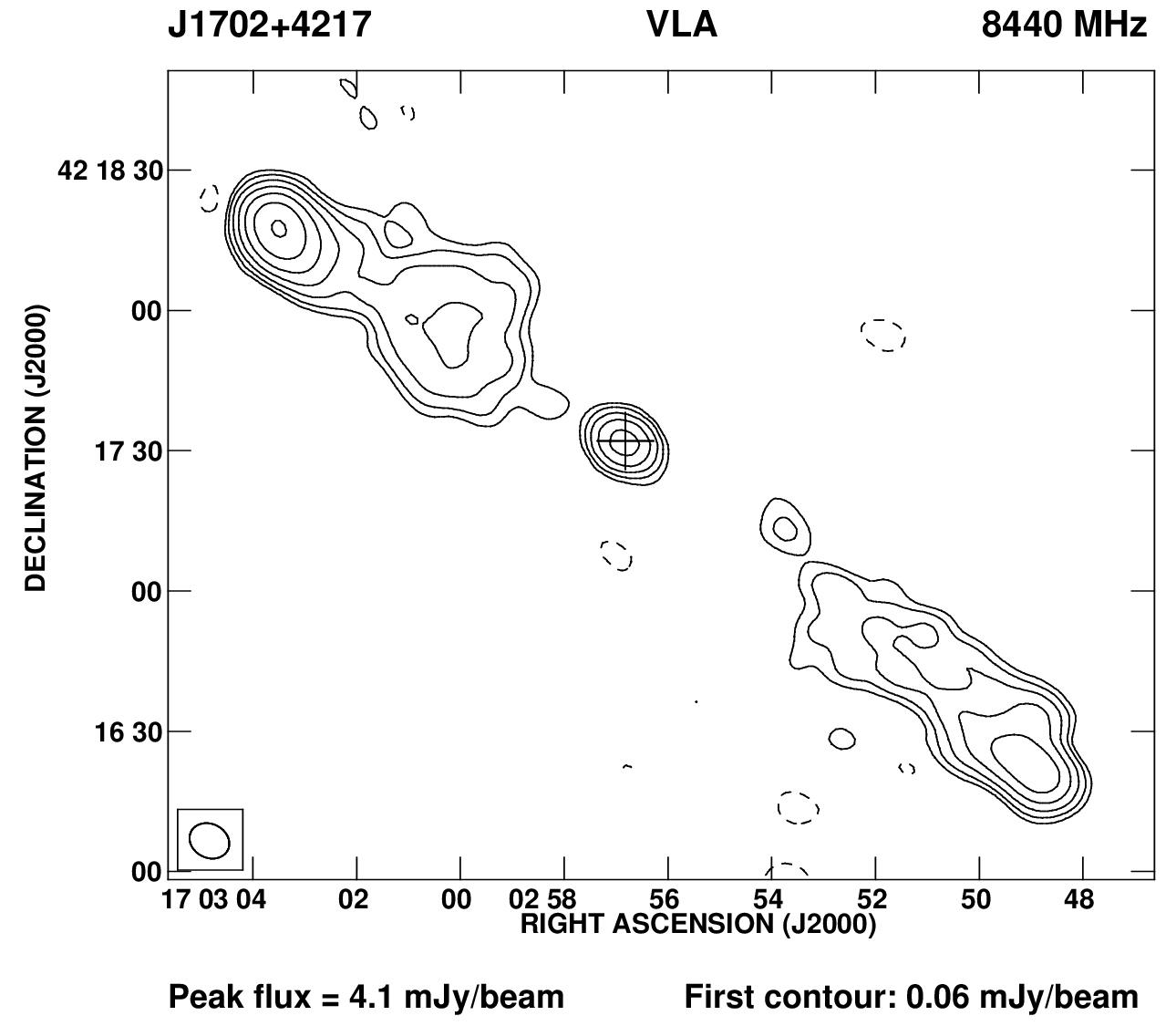,width=3.2in,angle=-90}
         }  
}
\contcaption{}
\end{figure*}

J1604+3731, 7C: The GMRT 610-MHz image  suggests a  
curved twin-jet structure on opposite sides of the core (Konar et al. 2004). 
The core is visible in all the GMRT images and in the 
VLA 4860-MHz and 9040 MHz images made from the archival data. 
The core has a straight steep spectrum with a spectral 
index of about 0.76$\pm$0.05, which may be due to either small-scale jets and/or lobes 
not resolved in the existing observations. This needs to be investigated. 

J1702+4217, 7C: The core has a flat-spectrum with a spectral index of
$\sim$0.2 between 600 and 5000 MHz, and the core flux density rising towards
higher frequencies.

J2312+1845, 3C457: The 334-MHz image of the source shows its large scale structure
with a barely detected core. The core is contaminated with the diffuse emission of the northern
lobe. The core spectral index is $\sim$0.1 between 1425 and 4866 MHz (see Konar et al. 2004).

\begin{table*}
\caption {Physical properties of the sources} 

\begin{tabular}{l c c l r r c l c c l}
\hline
Source &  Alt.  & Opt.& Redshift&LAS   &{\it l}&P$_{1.4}$ & f$_{\rm c}$ & R$_{\theta}$ & R$_{\rm S}$  & References \\
       & name   & Id. &         &      &       &          &             &              &              &     \\
       &        &     &         &$^{\prime\prime}$&kpc&W Hz$^{-1}$ &    &              &              &     \\
 (1)   & (2)    & (3) &  (4)    &  (5) &  (6)  &   (7)    &   (8)       &     (9)      &   (10)       & (11) \\
\hline                                                                                                  
J0720+2837&         &G & 0.2705 &370  &1520 & 25.02  & 0.06             &    1.37      &    1.06      & 1   \\
J0912+3510&         &G & 0.2489 &375  &1449 & 25.44  &$\lapp$0.009      &    1.66      &    1.82      & 1   \\
J0927+3510&         &G & 0.55*  &345  &2206 & 26.02  & 0.005            &    1.02      &    1.48      & 1   \\
J1155+4029&         &G & 0.53*  &229  &1437 & 26.55  & 0.14             &    2.00      &    0.15      & 1   \\
J1313+6937& DA 340  &G & 0.106  &388  & 745 & 25.60  & 0.01             &    1.30      &    0.68      & 2,4   \\ 
J1343+3758&         &G & 0.2267 &684  &2463 & 25.32  & 0.03             &    1.32      &    0.63      & 1,3  \\
J1604+3438&         &G & 0.2817 &200  & 846 & 25.53  & 0.04             &    1.09      &    1.20      & 1   \\
J1604+3731& 7C      &G & 0.814  &178  &1346 & 26.60  & 0.04             &    1.02      &    0.69      & 1,5,P   \\
J1702+4217& 7C      &G & 0.476  &196  &1160 & 26.20  & 0.03             &    1.33      &    0.69      & 5,P   \\
J2312+1845& 3C457   &G & 0.427  &190  &1056 & 27.11  & 0.006            &    1.05      &    0.75      & 2  \\
\hline
\end{tabular}

1: Machalski et al. (2006); 2: Konar et al. (2004); 3: Jamrozy et al. (2005); 
4: Saunders, Baldwin \& Warner (1987); 5: Cotter, Rawlings \& Saunders (1996); 
P: Present paper. \\
An asterisk in column 4 indicates an estimated redshift.
\end{table*}

\section{Discussion and results}
Some of the physical properties of the sources are listed in Table 3 which is
arranged as follows.
Columns 1 and 2: source name and an alternative name.  Column 3:
optical identification where G denotes a galaxy;
column 4: redshift; columns 5 and 6: the largest angular size in arcsec and 
the corresponding projected linear size in kpc; 
column 7: the luminosity at 1.4 GHz in logarithmic units of W Hz$^{-1}$;
column 8: the fraction of emission from the core, f$_c$,  at an emitted 
frequency of 8 GHz. For estimating f$_c$, core spectral indices have been estimated wherever 
possible, otherwise a value of 0 has been assumed. 
The separation ratio R$_{\theta}$, defined to be the separation of the farther
hotspot or lobe from the core or optical galaxy to the nearer one, and the flux density ratio,
R$_{\rm S}$ at an emitted frequency of 5 GHz in the same sense, are listed in 
columns 9 and 10 respectively.
References for the radio structure which contains a 5-GHz image and the redshift 
are listed in column 11.
For monochromatic luminosity at 1400 MHz emitted frequency, we have used NVSS images for the
total flux density and spectral indices between $\sim$600 and 5000 MHz for all sources except
J2312+1845 for which we have used the spectral index between $\sim$1400 and 5000 MHz. 

\subsection{Radiative losses}
For all the lobes listed in Table 2 we first determine the minimum energy magnetic field 
strength using the formalism of Miley (1980) by integrating the spectrum between 10 MHz and 100 GHz. 
We have repeated the calculations by integrating the spectrum from a frequency corresponding to a
minimum Lorentz factor, $\gamma_{\rm min}$$\sim$10 for the relativistic electrons to an upper
limit of 100 GHz, which corresponds to a Lorentz factor ranging from a few times 10$^4$ to 10$^5$ 
depending on the estimated magnetic field strength 
(see Hardcastle et al. 2004; Croston et al. 2005 and Appendix A).
These estimates are referred to as classical-1 and classical-2 respectively
in this paper.  The expressions we use for the more general case of a curved radio spectrum in 
the classical formalisms are described in Appendix A. 
We also estimate the magnetic field strength, B$_{\rm eq}$(rev), using the formalism of Beck \& Krause
(2005) which is an equipartition magnetic field.  Their formula (equation A18) has the parameter
{\bf K}$_{0}$ which is the ratio of the number density of protons to that of electrons in the 
energy range where losses are small. It is relevant to note that in this formalism particle energy
is dominated by the protons. Estimating {\bf K}$_{0}$=$(\frac{m_p}{m_e})^{\alpha}$ as given by
their equation (7) for $\alpha\approx \alpha_{\rm inj}$ which depends on the low-frequency spectral
index in the observed synchrotron spectrum, we can constrain the proton spectrum and hence 
estimate the revised magnetic field strength, B$_{\rm eq}$(rev).

We have estimated the magnetic field strengths for the extended lobes of all the sources in our
sample using the  different approaches, except for J0720+2837 where the spectral coverage is 
comparatively poor. We have
assumed either a cylindrical or spheroidal geometry and a filling factor of unity, and 
have estimated the sizes of the lobes from the lowest contours in the available low-frequency
images at either 330 or 605 MHz.  In some sources such as the the eastern lobe of J1343+3758 the 
emission appears quite compact while in others such as in J1702+4217 there are prominent 
bridges of emission. 
The magnetic field in nT (1T = 10$^{4}$ G) estimated from the different approaches as well
as the equivalent magnetic field of the cosmic microwave background radiation (CMBR) at the redshift
of the source, B$_{\rm iC}$=0.32(1 + z)$^2$ nT, are listed in Table 4, which is self explanatory.  
A comparison of the magnetic field estimates from the different approaches (Figure 2) shows
that the field strengths estimated using the Beck \& Krause and classical-2 formalisms are larger than 
those of classical-1 by factors of $\sim$3 and $\sim$2 respectively. An independent check of our classical-2
estimates was done by Martin Hardcastle using the formalism of Hardcastle et al. (2004) and the values
were found to be consistent.

\begin{figure}
\psfig{file=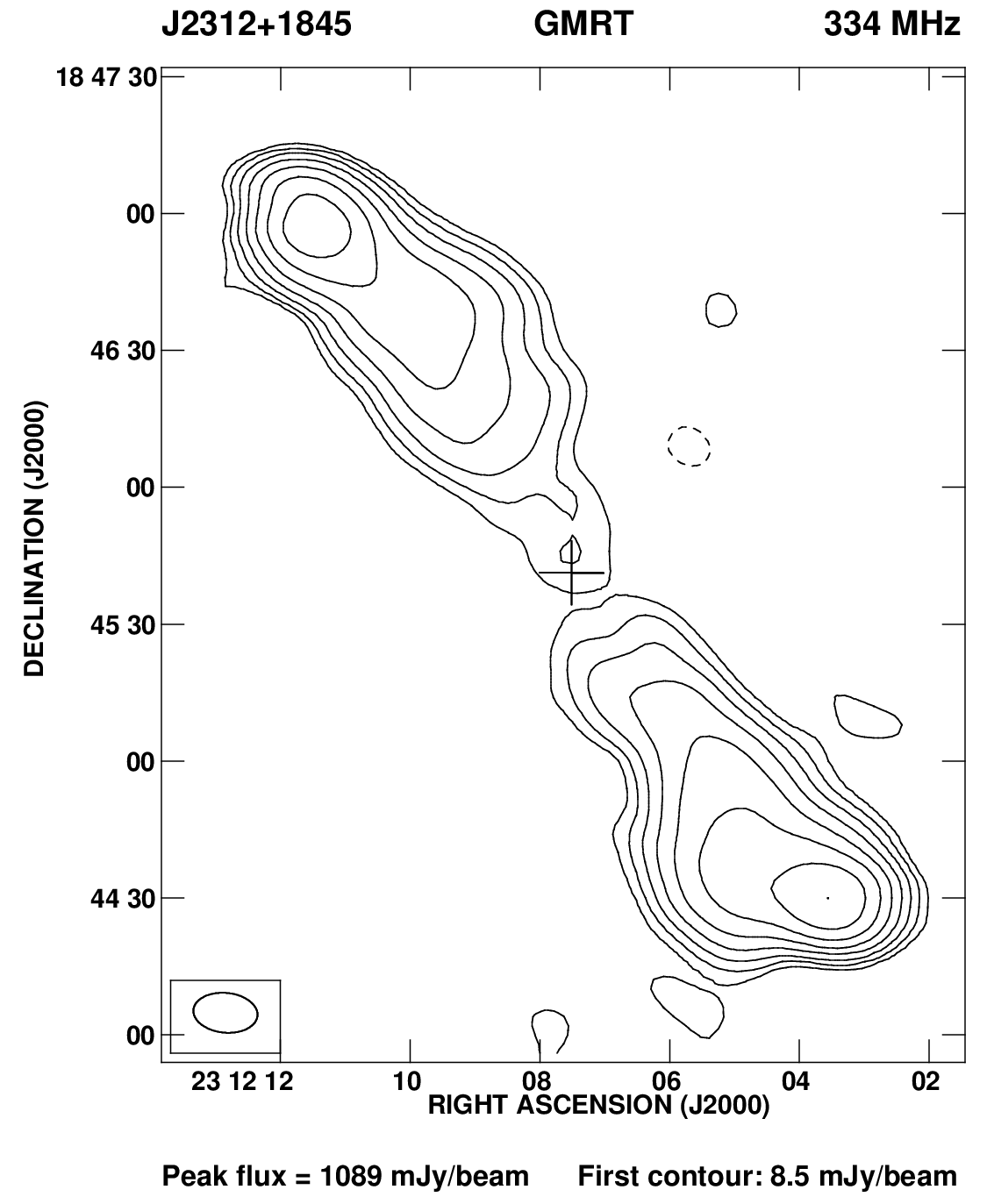,width=3.0in,angle=0}
\contcaption{}
\end{figure}

We examine the relative importance of synchrotron and inverse-Compton losses in the lobes of 
these radio sources. The ratio of the magnetic field strength estimated from the different 
approaches to that of the equivalent magnetic field of the CMBR for the different lobes are
shown in Fig. 3. It is seen clearly that the magnetic field estimated from the classical-1 approach 
is always less than B$_{\rm iC}$, suggesting that inverse-Compton losses are usually larger
than the synchrotron radiative losses for giant radio sources as suggested earlier 
(e.g. Gopal-Krishna, Wiita \& Saripalli 1989; Ishwara-Chandra \& Saikia 1999; Konar et al. 2004). 
The median value of the ratio, r$_{\rm B}$,
of the magnetic field estimate to B$_{\rm iC}$ is $\sim$0.34. However, for the field estimates using 
the classical-2 and Beck \& Krause formalisms the corresponding values of the ratio r$_{\rm B}$ are 
$\sim$0.51 and 0.91  respectively. In the latter case, synchrotron losses are more important in 
approximately half the cases. This could have a significant effect on the identification of GRSs.
For example, a combination of increased synchrotron losses combined with the inverse-Compton losses
could make it more difficult to detect the bridges of emission in high-redshift GRSs, thereby leading
to the classification of the hotspots and a possible core as unrelated radio sources.

\begin{table*}
\caption{Magnetic field estimates of the lobes}

\begin{tabular}{l c c c c c c c}
\hline
Source  &  Comp. &  \multicolumn{5}{c}{Magnetic field estimates} & Equivalent     \\
        &        &  JP: class-1    & JP: class-2   & KP: class-1   & KP: class-2   & Beck-Krause & CMB field \\  
        &        &    nT         &    nT         &   nT          &    nT         &   nT        &    nT     \\
 (1)    &  (2)   &   (3)         &    (4)        &   (5)         &    (6)        &   (7)       &    (8)    \\
\hline
J0912+3510  & N  & 0.11$\pm$0.01 & 0.16$\pm$0.02 & 0.11$\pm$0.01 & 0.16$\pm$0.02 & 0.34$\pm$0.05 & 0.50  \\
            & S  & 0.13$\pm$0.01 & 0.21$\pm$0.02 & 0.13$\pm$0.01 & 0.21$\pm$0.02 & 0.39$\pm$0.05 &       \\
J0927+3510  & NW & 0.13$\pm$0.01 & 0.27$\pm$0.03 & 0.13$\pm$0.01 & 0.27$\pm$0.03 & 0.38$\pm$0.05 & 0.77  \\
            & SE & 0.14$\pm$0.01 & 0.25$\pm$0.02 & 0.14$\pm$0.01 & 0.25$\pm$0.02 & 0.35$\pm$0.04 &       \\
J1155+4029  & NE & 0.50$\pm$0.05 & 1.15$\pm$0.11 & 0.50$\pm$0.05 & 1.13$\pm$0.11 & 1.52$\pm$0.18 & 0.75  \\
            & SW & 0.32$\pm$0.03 & 0.70$\pm$0.07 & 0.31$\pm$0.03 & 0.68$\pm$0.07 & 0.99$\pm$0.12 &       \\
J1313+6937  & NW & 0.19$\pm$0.02 & 0.29$\pm$0.03 & 0.19$\pm$0.02 & 0.29$\pm$0.03 & 0.60$\pm$0.08 & 0.39  \\
            & SE & 0.15$\pm$0.01 & 0.23$\pm$0.02 & 0.15$\pm$0.01 & 0.23$\pm$0.02 & 0.48$\pm$0.06 &       \\
J1343+3758  & SW & 0.13$\pm$0.01 & 0.23$\pm$0.02 & 0.13$\pm$0.01 & 0.23$\pm$0.02 & 0.40$\pm$0.05 & 0.48  \\
            & NE & 0.20$\pm$0.02 & 0.33$\pm$0.03 & 0.20$\pm$0.02 & 0.32$\pm$0.03 & 0.60$\pm$0.08 &       \\
J1604+3438  & W  & 0.17$\pm$0.02 & 0.24$\pm$0.02 & 0.17$\pm$0.02 & 0.24$\pm$0.02 & 0.52$\pm$0.07 & 0.53  \\
            & E  & 0.20$\pm$0.02 & 0.28$\pm$0.03 & 0.20$\pm$0.02 & 0.28$\pm$0.03 & 0.60$\pm$0.08 &       \\
J1604+3731  & N  & 0.26$\pm$0.03 & 0.52$\pm$0.05 & 0.26$\pm$0.03 & 0.51$\pm$0.05 & 0.58$\pm$0.07 & 1.05  \\
            & S  & 0.37$\pm$0.04 & 0.73$\pm$0.07 & 0.37$\pm$0.04 & 0.72$\pm$0.07 & 0.83$\pm$0.10 &       \\
J1702+4217  & NE & 0.20$\pm$0.02 & 0.30$\pm$0.03 & 0.20$\pm$0.02 & 0.30$\pm$0.03 & 0.48$\pm$0.06 & 0.70  \\
            & SW & 0.20$\pm$0.02 & 0.30$\pm$0.03 & 0.20$\pm$0.02 & 0.30$\pm$0.03 & 0.47$\pm$0.06 &       \\
J2312+1845  & NE & 0.40$\pm$0.04 & 0.86$\pm$0.08 & 0.40$\pm$0.04 & 0.85$\pm$0.08 & 1.23$\pm$0.15 & 0.65  \\
            & SW & 0.43$\pm$0.04 & 0.91$\pm$0.09 & 0.43$\pm$0.04 & 0.90$\pm$0.09 & 1.32$\pm$0.16 &       \\
\hline
\end{tabular}
\end{table*}

\begin{figure}
\vbox{
     \psfig{file=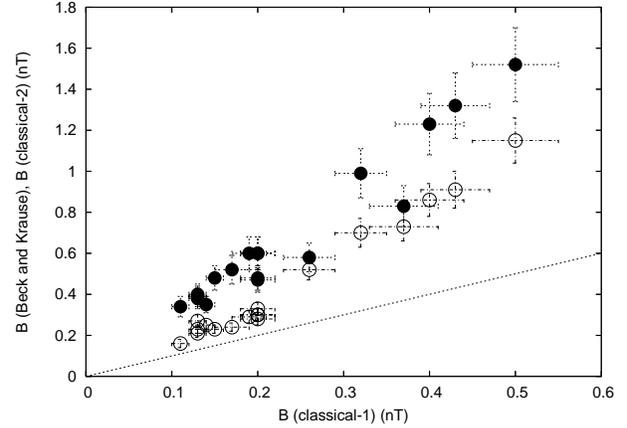,width=3.25in,angle=-90}
    }
\caption{The magnetic field strength estimated using the formalisms of Beck \& Krause (2005) shown
by filled circles and classical-2 (see text and Croston et al. 2005) shown by open circles are plotted 
against the classical-1 estimates (see text and Miley 1980). The dotted line represents field strengths
estimated using the Beck \& Krause and classical-2 formalisms being equal to the classical-1 estimates.
}
\end{figure}

\begin{figure}
\vbox{
     \psfig{file=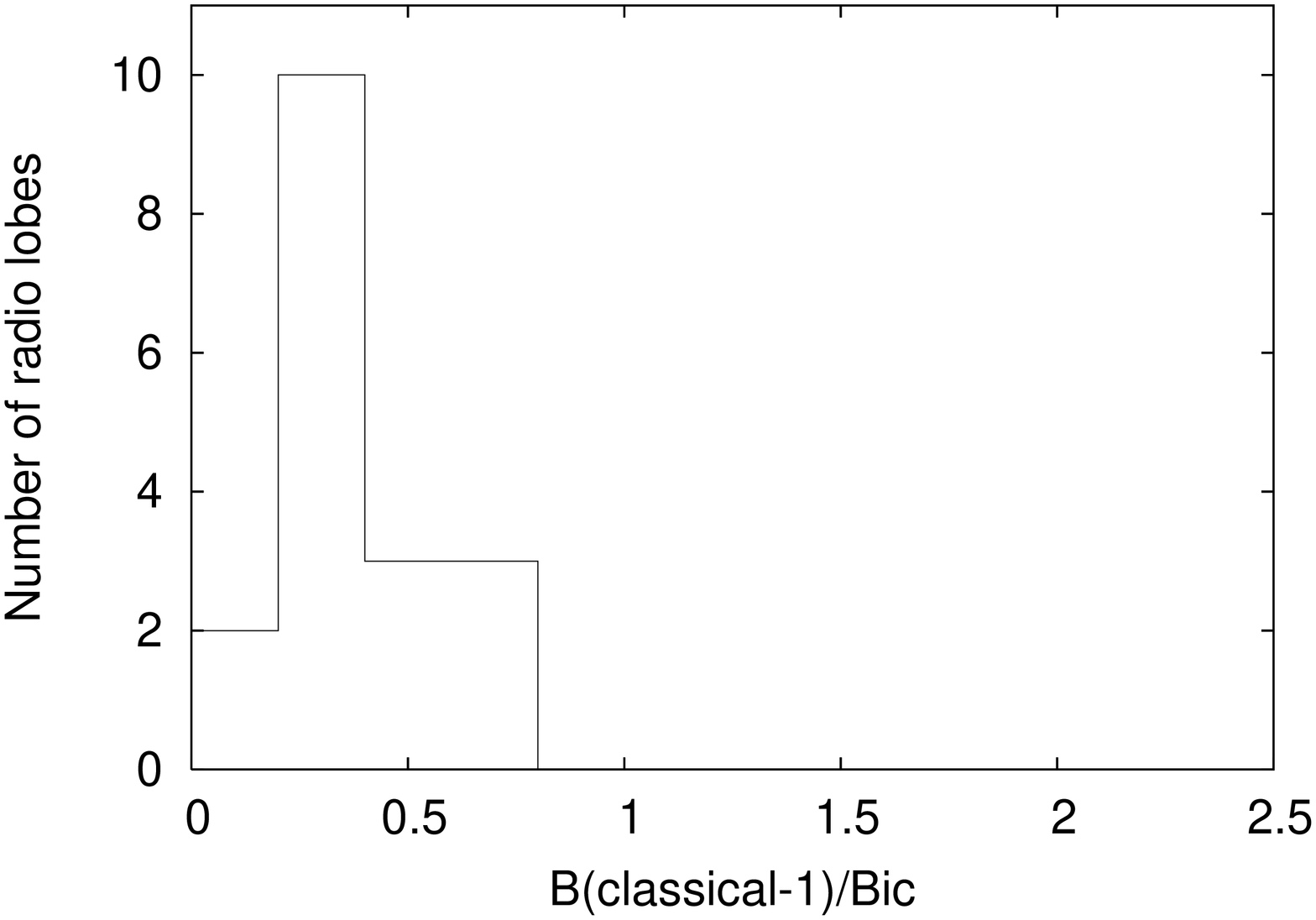,width=2.5in,angle=-90}
     \psfig{file=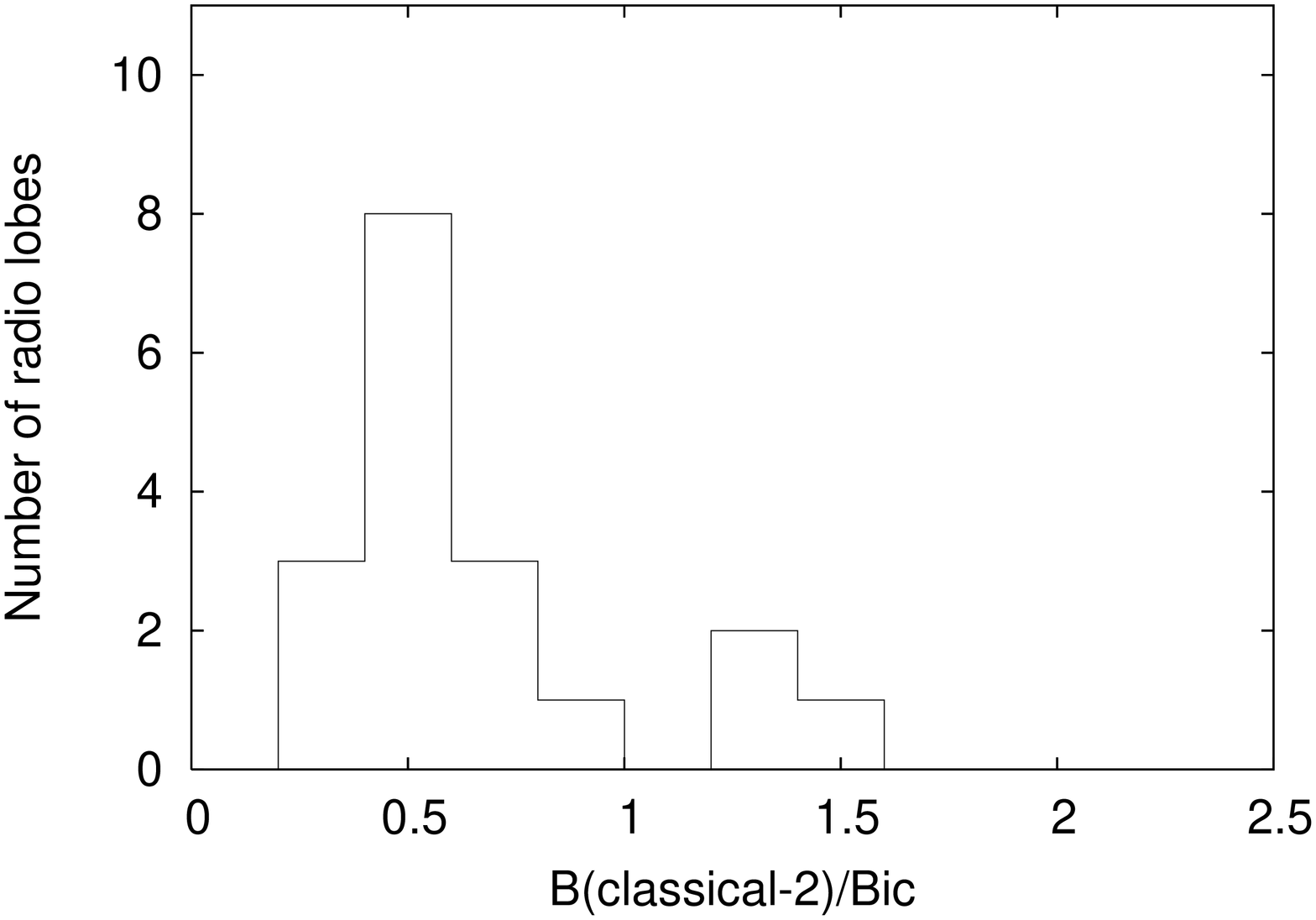,width=2.5in,angle=-90}
     \psfig{file=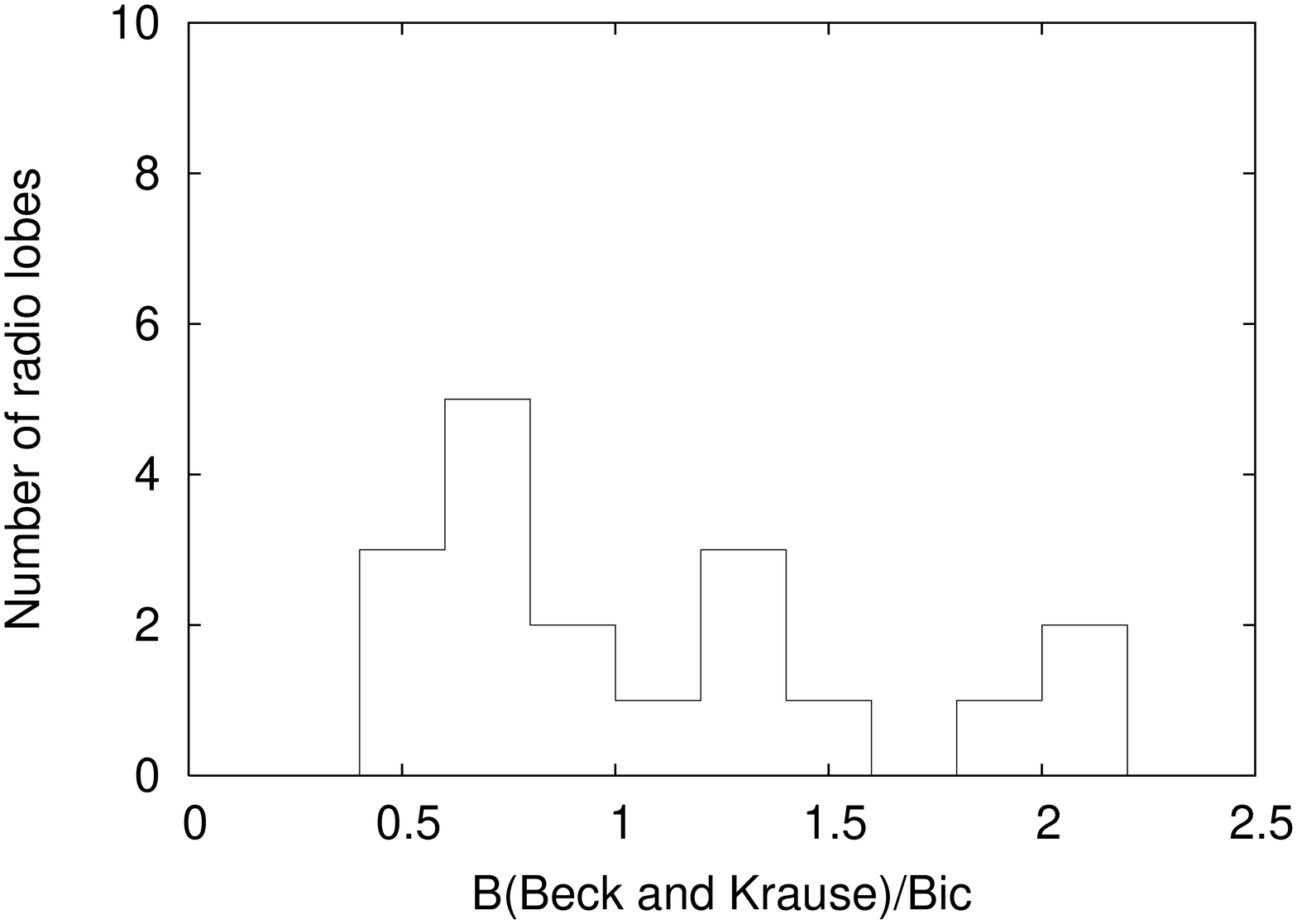,width=2.5in,angle=-90}
    }
\caption{The distributions of the ratios of the magnetic field strength estimated using the classical-1, 
classical-2 and Beck \& Krause (2005) formalisms and the equivalent magnetic field, B$_{\rm iC}$, of
the CMBR, shown in the upper, middle and lower panels respectively.}
\end{figure}

\subsection{Core properties}
The fraction of emission from the core at an emitted frequency of 8 GHz, f$_{\rm c}$, ranges from 
$\sim$0.005 to 0.14, with a median value of $\sim$0.03. The values of f$_{\rm c}$ are
usually comparable to the other sources of similar radio luminosity. The sources with prominent
cores (f$_{\rm c}$$\gapp$0.04) are J0720+2837, J1155+4029, J1604+3438 and J1604+3731. Of these four 
sources J0720+2837 has a steep two-point spectral index, J1155+4029 has a 
giga-Hertz peaked spectrum (GPS) core, J1604+3438 has a flat-spectrum core while 
J1604+3731 has a steep spectrum core from $\sim$300 to 9000 MHz (see Section 3.1). The flux 
densities of the cores of 
J1155+4029 and J1604+3731 where multiple measurements are available are listed in Tables 5 and 6
respectively along with the approximate resolution and the epoch of observations, and their
spectra are presented in Fig. 4. For sources
where contamination of core emission by diffuse extended emission at the lower frequencies seem
significant, the source has been remapped with a lower uv-cutoff of 5 k$\lambda$ to minimise
contamination by the extended emission.  
It would be interesting to determine from mas-scale resolution observations whether the
core of J1155+4029 is resolved into a pc-scale double. It is worth mentioning that the GPS core of the GRS, J1247+6723,
is resolved into a small-scale double with a size of $\sim$14 pc, suggesting recurrent activity
(Marecki et al. 2003; Saikia, Konar \& Kulkarni 2006). It is also relevant to note that the 
core does not exhibit any evidence of significant 
variability as in the case of other GPS sources (e.g. O'Dea 1998 for a review). For example the 
core flux densities from the FIRST  survey at 
1400 MHz and our GMRT 1258-MHz image are similar even though they are temporally far apart. 

\begin{figure}
\vbox{
     \psfig{file=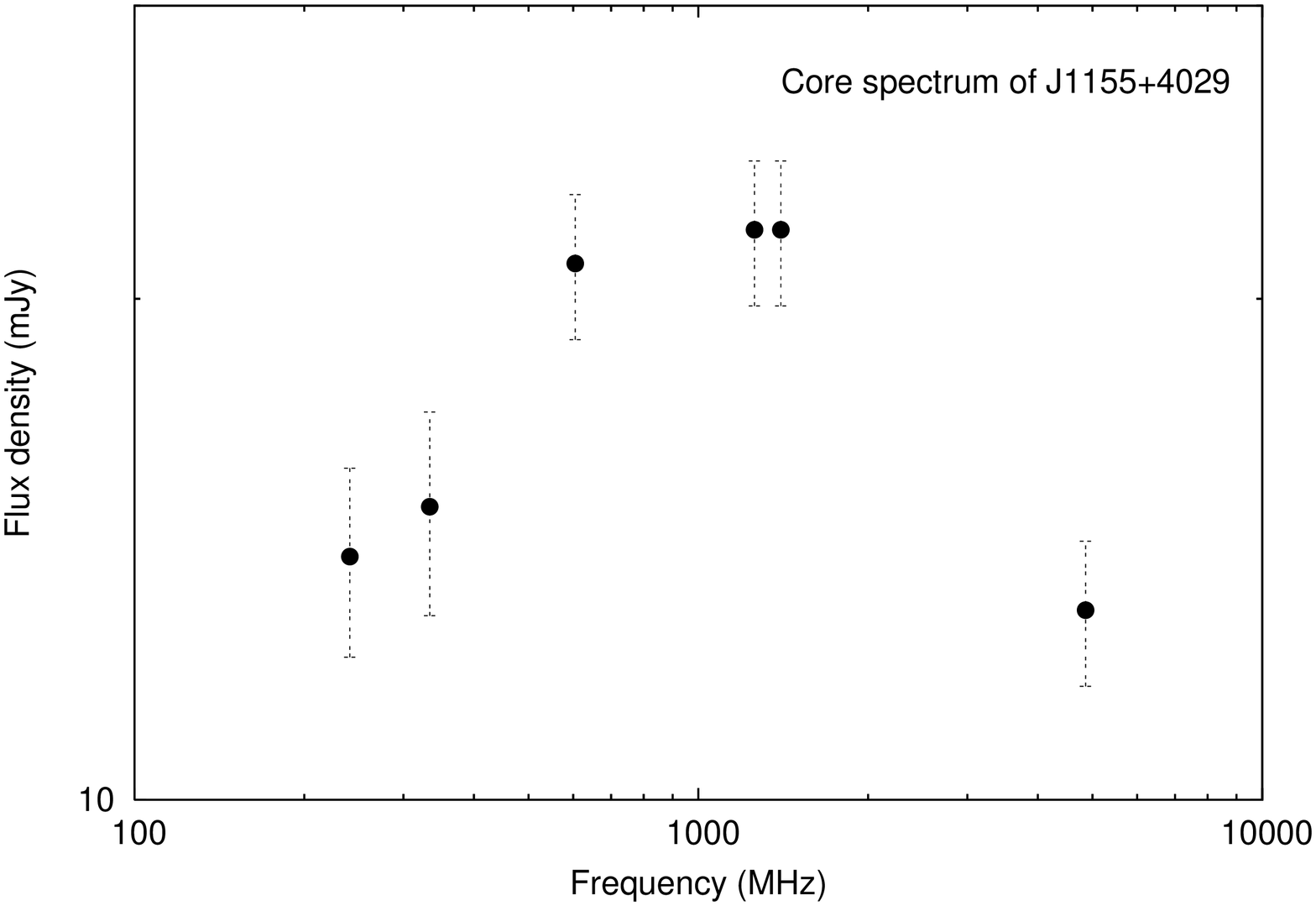,width=3.25in,angle=-90}
     \psfig{file=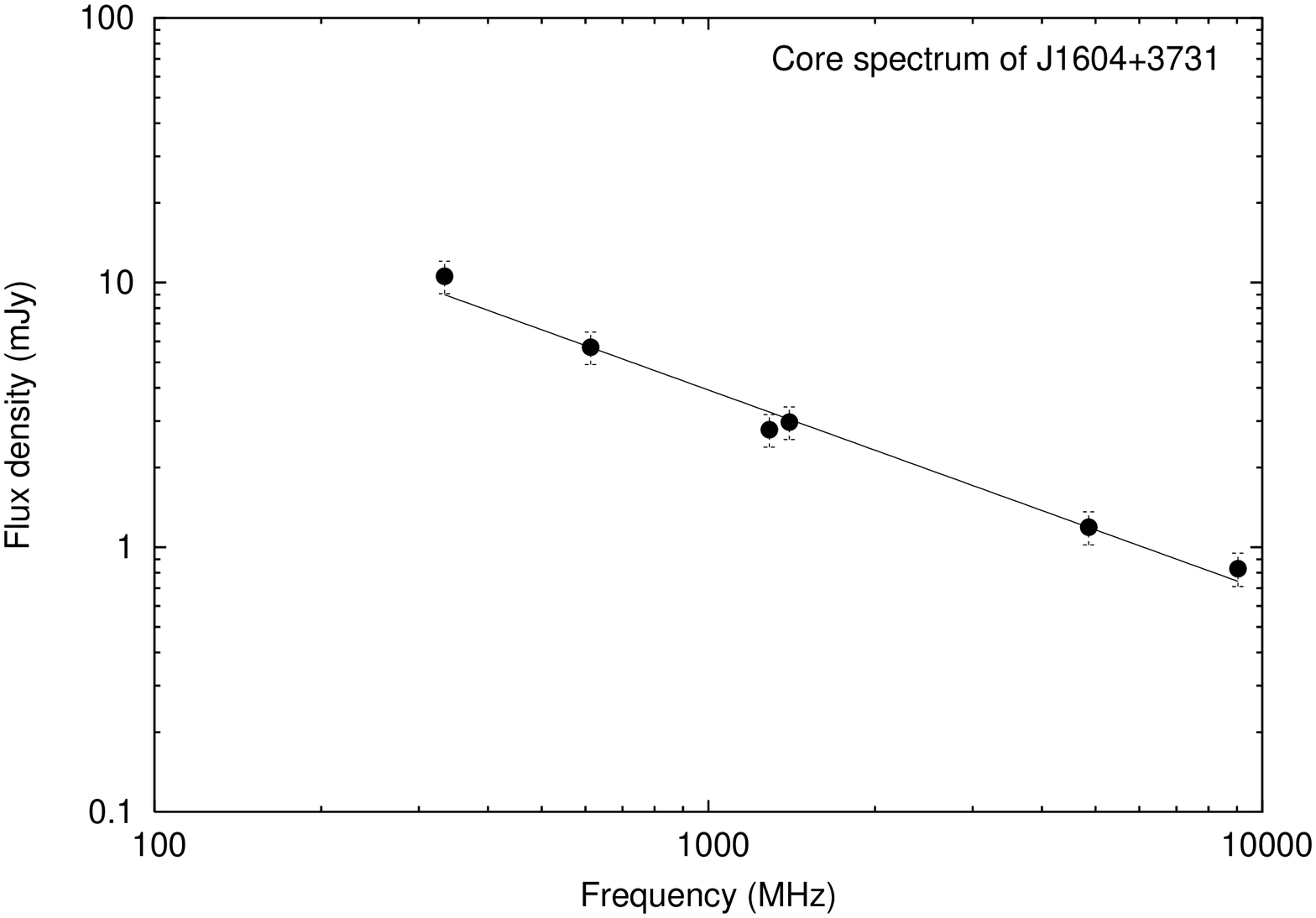,width=3.25in,angle=-90}
    }
\caption{Upper panel: core spectrum of J1155+4029; lower panel: a least-square fit of a power-law to the core
spectrum of J1604+3731.}
\end{figure}

\subsection{Environment and morphology of GRSs}
The lobes of the GRSs lie well beyond the extent of the parent host galaxies and can be used to
probe the environment on these scales which could affect the structure and symmetry parameters 
of these objects (e.g Subrahmanyan et al. 1996). Although GRSs often occur in regions of low
galaxy density, Subrahmanyan et al. noted that two of the sources in their sample, namely 
B1545$-$321 and B2356$-$611, which have continuous bridges from the hotspots to the core
show significant excesses in galaxy counts. Ishwara-Chandra \& Saikia (1999) and Schoenmakers
et al. (2000) noted that the GRSs tended to be marginally more asymmetric than smaller sources
of similar luminosity, excluding the compact steep spectrum sources. For example, considering
the higher luminosity objects Ishwara-Chandra \& Saikia noted that the
median value of the separation ratio is $\sim$1.39 compared with $\sim$1.19 for 3CR galaxies
of similar luminosity but smaller sizes. The median value for these ten sources in  our sample
is $\sim$1.31 (see Fig. 5, upper panel), consistent with the earlier estimate although the statistical uncertainties
are now larger because of the small sample size. The distribution does not show a significant
peak towards smaller values. For a source inclined at $\gapp$45$^\circ$ to the line of sight,
the expected separation ratio is only $\sim$1.15 for a hotspot advance speed of $\sim$0.1c
(see Scheuer 1995).  A plot of the separation ratio against the fraction of emission 
from the core (Fig. 5, upper panel), which is often used as a statistical indicator of orientation of the source axis
to the line of sight, shows no significant correlation suggesting that large-scale environmental
asymmetries play a significant role. It is worth noting that although J1155+4029, which has the
most prominent core, and has the highest value of the separation ratio, the nearer component
is brighter by a factor of $\sim$6 (Fig. 5, lower panel), demonstrating that density asymmetries  
are likely to again play a significant role in the observed asymmetries. 
It is also relevant to note that
the radio source J1155+4029 which has the highest value of f$_{\rm c}$=0.14
has a GPS core; it is important to determine its structure with mas resolution and investigate 
whether it might be a compact double with a rather weak core. 

Although flux density
ratios would be affected by relative contributions of the hotspot and backflow emission which have 
different velocities, in addition to effects of evolution of individual components with age,
it can be seen in Fig. 5 (middle and lower panels) that in 6 of the 10 sources the nearer 
component is brighter as is expected if their
jets are going through a denser medium (e.g. Jeyakumar et al. 2005). In spite of the difficulties
noted above, weak trends for the components on the jet side in quasars to be stronger have been
reported (e.g Garrington, Conway \& Leahy 1991), illustrating that relativistic effects also play 
a role. However, in these giant radio sources, environmental effects seem to dominate, with the flux
density ratio too exhibiting no significant dependence on core prominence (Fig. 5, middle panel). It would be 
interesting to explore the fields of these sources both via optical galaxy counts and deep x-ray observations,
and examine their relationships with the observed structures.
  
\begin{figure}
\vbox{
     \psfig{file=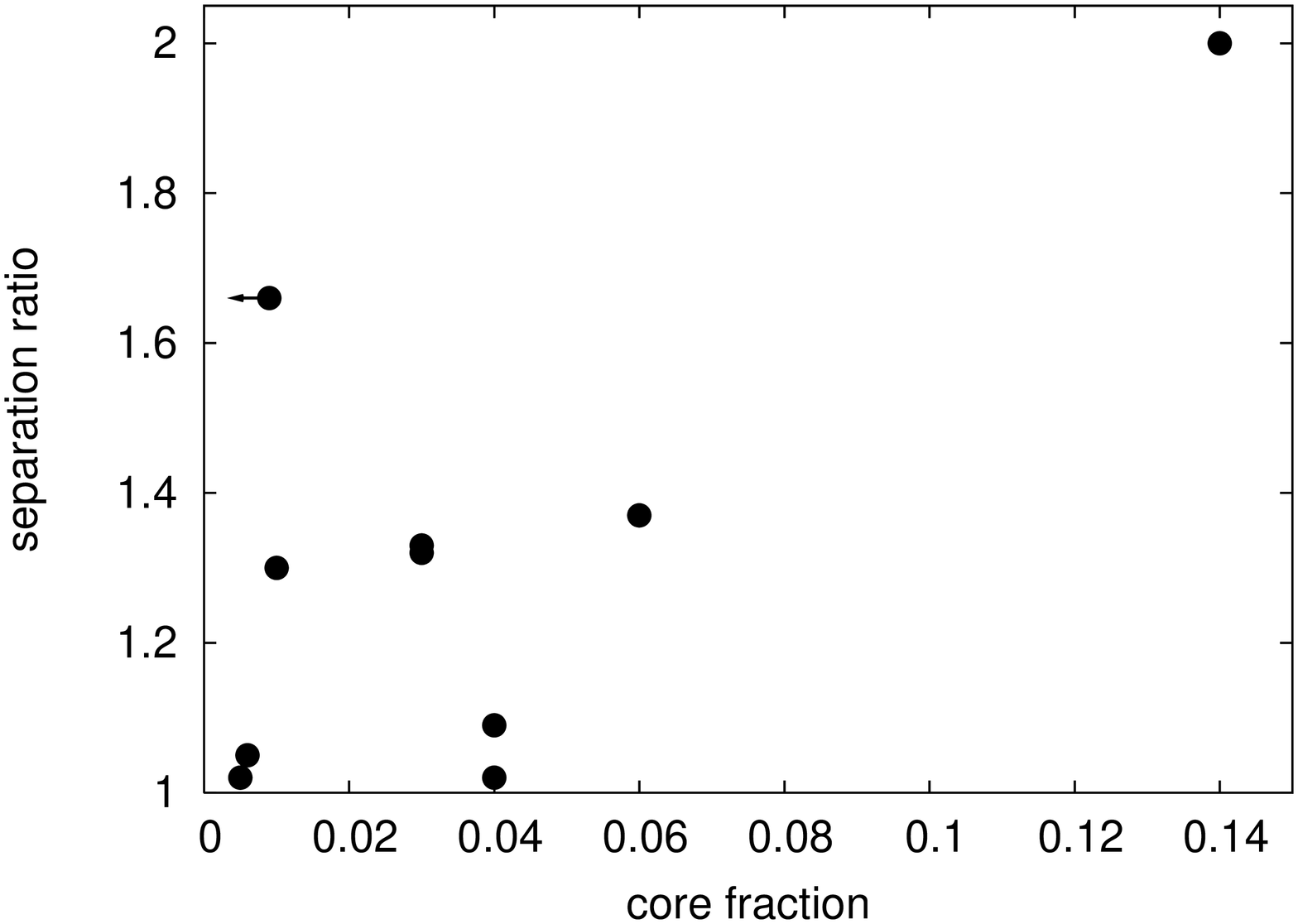,width=3.0in,angle=-90}
     \psfig{file=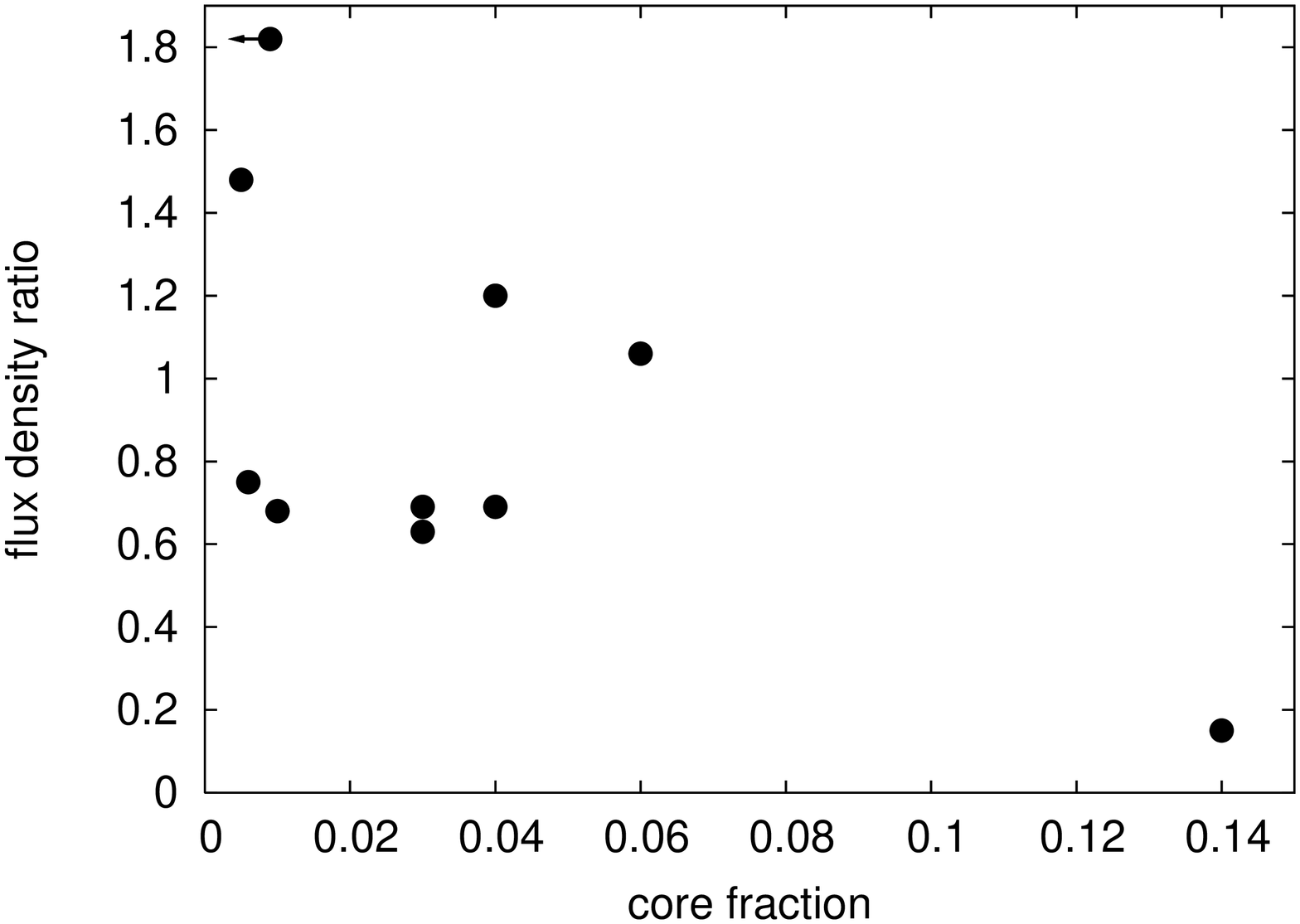,width=3.0in,angle=-90}
     \psfig{file=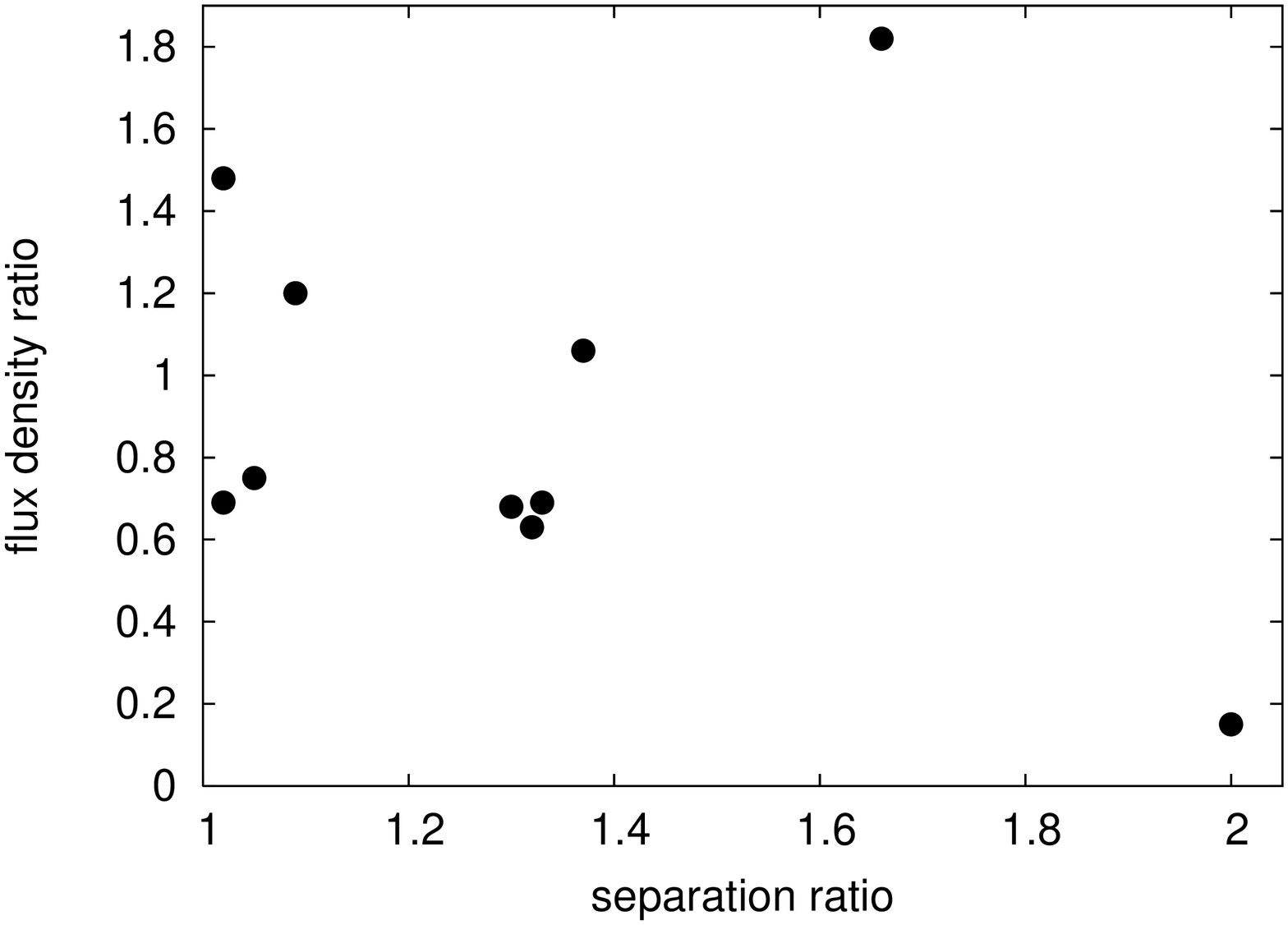,width=3.0in,angle=-90}
    }
\caption{The separation ratios and flux density ratios are plotted against the fraction of emission from the core
at an emitted frequency of 8 GHz in the upper and middle panels respectively. In the lower panel the flux density
ratio is plotted against the separation ratio.}
\end{figure}

\begin{table}
\caption{Core flux densities of J1155+4029}
\begin{tabular}{l r l r r r}
\hline
 Teles-    &  resn.             &     Date  & Freq.   &  S    & Ref.   \\
 cope      &  $^{\prime\prime}$ &           & MHz     &  mJy  &        \\
  (1)      &     (2)            &   (3)     &  (4)    &  (5)  & (6)    \\
\hline

GMRT       &  9.9               &2005 Mar 16 & 241     &       14 & 1  \\
GMRT       &  9.3               &2005 Dec 25 & 334     &       15 & 1  \\
GMRT       &  6.1               &2005 Mar 16 & 605     &       21 & 1  \\
GMRT       &  9.9               &2005 Nov 30 &1258     &       22 & 1  \\
VLA-B      &  5.4               & FIRST      &1400     &       22 & 2  \\
VLA-D      & 14.7               &2004 Jul 31 &4860     &       13 & 1  \\
\hline
\end{tabular}
~~~~~~~~~~~~~~1: Present paper; 2: FIRST
\end{table}

\begin{table}
\caption{Core flux densities of J1604+3731}
\begin{tabular}{l r l r r r}
\hline
 Teles-    &  resn.             &     Date  & Freq.   &  S    & Ref.   \\
 cope      &  $^{\prime\prime}$ &           & MHz     &  mJy  &        \\
  (1)      &     (2)            &   (3)     &  (4)    &  (5)  & (6)    \\
\hline

GMRT       &  7.2              &2006 Mar 09& 334     & 10.6 &  1    \\
GMRT       &  5.1              &2004 Jan 01& 613     &  5.7 &  1    \\ 
GMRT       &  3.5              &2006 Jan 01&1289     &  2.8 &  1    \\
VLA-BnD    &  5.4              &     FIRST &1400     &  3.0 &  2    \\
VLA-D      & 10.6              &1993 Dec 23&4860     &  1.2 &  1    \\ 
VLA-D      &  6.6              &1993 Dec 23&9040     &  0.8 &  1    \\
\hline
\end{tabular}
~~~~~~~~~~~~~~1: Present paper; 2: FIRST
\end{table}

\section{Concluding remarks}
We have presented multifrequency radio images of a selected sample of ten
giant radio galaxies, largely at low radio frequencies with the GMRT. We have
also presented a few high-frequency images made from VLA data. We
have listed the flux densites of components for the images presented here as
well as for high-frequency images of these sources which have been published 
earlier by us (MJZK; Konar et al. 2004). 

From these multifrequency observations
of the lobes we estimate the magnetic field strengths using three
different approaches, namely estimating the minimum energy field by integrating
from 10 MHz to 100 GHz which has been referred to here as classical-1 (Miley 1980), 
integrating from a minimum frequency corresponding to
$\gamma_{\rm min}$=10 at the estimated magnetic field to a maximum frequency of 100 GHz,
called classical-2 (see Hardcastle et al. 2004; Croston et al. 2005), and lastly using the
formalism of Beck \& Krause (2005). We show that on the average the magnetic field strengths
estimated using the Beck \& Krause and classical-2 formalisms are larger than the classical-1 
values by factors of $\sim$3 and $\sim$2 respectively. 

The inverse-Compton losses dominate over synchrotron losses when estimates of the
classical  minimum energy magnetic field are used, consistent with earlier studies.
However, this is often not true if the magnetic fields are close to the values estimated
using the formalism of Beck \& Krause (2005). In the latter case, synchrotron losses
are more important in nearly half the cases. 
The increased synchrotron losses combined with the inverse-Compton losses
would make it more difficult to detect the bridges of emission in high-redshift GRSs, 
thereby leading to the classification of the hotspots and a possible core as unrelated 
radio sources. Although  for a sample of radio galaxies
with x-ray emission from at least one of the lobes which can be attributed to inverse Compton
scattering with the CMBR photons, the estimated magnetic field ranges from $\sim$0.3 to 1.3 times the 
minimum energy field with a peak around 0.7 times this field (Croston et al. 2005), 
none of the radio lobes in clusters of galaxies studied by Dunn, Fabian \& Taylor (2005)
has equipartition between the relativistic particles and the magnetic field. Clearly more
work is required in this area. 

We also examine the spectral indices of the cores and any evidence of recurrent activity 
in these sources. In one of the sources, J1155+4029, the core has a GPS spectrum, while in
two others, J0720+2837 and J1604+3731, the cores appear to have a steep spectrum. It would
be interesting to determine the milliarcsec-scale structure of the core in J1155+4029 as
well as J0720+2837 and J1604+3731 and thereby examine whether these show evidence of being
a double double radio galaxy. 

We probe the environment using the symmetry parameters of these sources. Approximately
half the sources are more asymmetric in their separation ratio than would be expected
for a galaxy inclined at $\gapp$45$^\circ$ to the line of sight with the hotspots moving
outwards with a velocity of $\sim$0.1c. Also the nearer lobe is brighter in six of the
sources, the asymmetry being  most pronounced in the case of J1155+4029 which has the 
largest value of separation ratio and a GPS core. These trends suggest that
the environments of these sources are often asymmetric on scales of $\sim$1 Mpc,
consistent with earlier studies.
  
\section*{Acknowledgments}
We thank Martin Hardcastle for his advice and comments and for making an independent 
check of our classical-2 estimates using the formalism of Hardcastle et al.
(2004). We also thank Rainer Beck, Gopal-Krishna, Martin Krause, Vasant Kulkarni, 
Paul Wiita and an anonymous referee for their comments on  the manuscript.
The Giant Metrewave Radio Telescope is a national facility operated by the National Centre 
for Radio Astrophysics of the Tata Institute of Fundamental Research. We thank the staff
for help with the observations.  The National Radio Astronomy Observatory  is a
facility of the National Science Foundation operated under co-operative
agreement by Associated Universities Inc. We thank the VLA staff for easy access
to the archival data base.  This research has made use of the NASA/IPAC extragalactic database (NED)
which is operated by the Jet Propulsion Laboratory, Caltech, under contract
with the National Aeronautics and Space Administration.  We thank numerous contributors
to the GNU/Linux group.

{}

\appendix

\section[]{Estimating classical equipartition magnetic field}

In this paper we estimate the magnetic field using the revised equipartition method (Beck \& Krause 2005) 
as well as the widely used classical minimum energy approach summarised by Miley (1980) and also 
a variant of this method which is described below. The latter two approaches are referred to as
classical-1 and classical-2 in this paper.  Since the expressions in the literature
for the classical-1 approach are for a power law spectrum, and many of the GRSs show evidence
of curvature in the integrated spectrum, we estimate the magnetic field by numerically integrating over
the non power-law spectrum. 

To derive a general formula for the minimum energy magnetic field for non power-law spectra we
proceed as follows. The equations used are summarised here. The critical frequency at which an electron 
of energy $E$ radiates most of its energy is given by 
\begin{equation}
\nu=C_1 B(sin\phi)E^2 
\label{critical_freq}
\end{equation} 
where $C_1= 6.266\times 10^{18}$ in cgs units, and is a  constant (Pacholczyk 1970). 
The energy loss of a single electron is given by 
\begin{equation}
-\frac{dE}{dt}=C_2B^2(sin^2\phi)E^2
\label{sing.elect.lum}
\end{equation}
where $C_2= 2.368\times 10^{-3}$ in cgs units, is also a constant (Pacholczyk 1970)\\

Let us assume that the number of electrons between $E$ and $E+dE$ is given by $N(E)dE$. They will
radiate via synchrotron emission between $\nu$ and $\nu+d\nu$, where $\nu$ and $E$ are related by 
the equation (\ref{critical_freq}). The monochromatic luminosity, $L(\nu)$, is given by 
\begin{equation}
L(\nu)d\nu = N(E)dE \times (-\frac{dE}{dt})
\label{lnu.dnu}
\end{equation}

Substituting the value of $-\frac{dE}{dt}$ from equation (\ref{sing.elect.lum}) we get 
\begin{equation}
N(E)dE = \frac{C_1}{C_2}\frac{1}{Bsin\phi}\frac{L(\nu)}{\nu}d\nu
\label{exprn.4.ne}
\end{equation}

The total kinetic energy of the electrons is given by
\begin{equation}
U_e=\int E N(E)dE = \frac{1}{B^{3/2}sin^{3/2}\phi}C_4\int \frac{L(\nu)}{\sqrt{\nu}}d\nu 
\label{total.elect.energy1}
\end{equation}
$E$ has been substituted from equation (\ref{critical_freq}) and $C_4$=$\frac{\sqrt{C_1}}{C_2}$
is $1.05709\times10^{12}$ in cgs units. 

The above expression can be re-written as 
\begin{equation}
U_e= \frac{A}{B^{3/2}sin^{3/2}\phi}
\label{total.elect.energy2}
\end{equation}
where
\begin{equation}
A=C_4\int \frac{L(\nu)}{\sqrt{\nu}}d\nu
\label{Integral}
\end{equation}

Assuming the proton to electron energy density ratio to be $\kappa$ we get the total energy
expression as
\begin{equation}
U_{tot}=(1+\kappa)U_e + U_B = \frac{(1+\kappa)A}{B^{3/2}sin^{3/2}\phi}  + V\frac{B^2}{8\pi}
\label{total.energy}
\end{equation}

Differentiating $U_{tot}$ with respect to $B$ and equating $\frac{dU_{tot}}{dB}$ to zero
we get the minimum energy expression 
\begin{equation}
B_{min} = (\frac{6\pi A(1+\kappa)}{V sin^{3/2}\phi})^{2/7}
\label{expren.4.B_min1}
\end{equation}
where $A$ is given by equation (\ref{Integral}), while
$V$ and $L(\nu)$ can be written as follows. \\
$V=\eta f_s \frac{d^2_A \theta^{\prime\prime}_x \theta^{\prime\prime}_y {\it s} }{(206265)^2}$,
and $L(\nu)=4\pi d^2_L S(\nu)$
Here, $\eta$ is the filling factor, $f_s$ is a shape factor for the volume of the emission region, which 
assumes a value of $\frac{\pi}{4}$ for a cylinder and $\frac{\pi}{6}$ for an ellipsoid, $d_A$ and
$d_L$ are angular-diameter and luminosity distances respectively, $\theta^{\prime\prime}_x$ and
$\theta^{\prime\prime}_y$ are the two projected dimensions of the emission region, {\it s} is 
the depth of the emission region.  

So far, all $\nu$ and $S(\nu)$ values are in emitter's frame. Converting them into the quantities of 
observer's frame and expressing the quantities in practical units we get 

\begin{equation}
B_{min}(\mu G) = 10^6(1+z)[\frac{3.26604\times10^{-32}}{sin^{3/2}\phi} 
\frac{(1+\kappa) A^{\prime} }{\eta f_s \theta^{\prime\prime}_x \theta^{\prime\prime}_y {\it s(kpc)}}]^{2/7}
\label{expren.4.B_min2}
\end{equation}
Here $A^{\prime}=C_4\int^{\nu_2^o}_{\nu_1^o}\frac{S^o_{mJy}(\nu^o)}{\sqrt{\nu^o_{MHz}}}d\nu^o_{MHz}$.
The superscript `o' indicates that the quantities are in the observer's frame. The integration 
should be carried out within a frequency interval which is compatible with the emitter's frame frequency 
interval set by us.  $(2/3)^{3/4}$ has been substituted for $sin^{3/2}\phi$ in the above equation to
estimate $B_{min}$, as the average value of $sin^2\phi = \frac{2}{3}$.

As a first step, we
fit the observed flux density measurements with the spectral ageing models of Jaffe \& Perola
(1973, hereinafter referred to as JP) as well as Kardashev (1962) and Pacholczyk (1970),
hereinafter referred as KP, with the help of the $\sc SYNAGE$ package (Murgia 1996), and
then extrapolate to very low and high frequencies. 
For the integration limits, we first use the range 10 MHz $-$ 100 GHz in the emitter's frame for
the classical-1 approach. For classical-2
we set the integration limit such that the lower limit corresponds to a Lorentz factor, $\gamma_{min}$ of
$\sim$10 while the upper limit corresponds to 100 GHz, as in classical-1 (see Myers \& Spangler 1985;
Hardcastle et al. 2004; Croston et al. 2005). Since, we don't
have an independent estimate of the magnetic field
to start with, we assume a lower limit, say, 10 MHz to start with, keeping the upper limit
fixed at 100 GHz. Then we calculate the magnetic field and check what value of $\gamma_{min}$
does the lower limit of frequency correspond to using the relation
$\nu=C_1Bsin\phi (\gamma -1)^2$m$_{\rm e}$c$^2$,
where m$_{\rm e}$ is the rest mass of the electron and c represents the velocity of light.
Then we vary the lower limit iteratively till we get $\gamma_{min}$ close to 10.

\end{document}